\documentstyle[epsfig]{mn}

\newif\ifAMStwofonts
\AMStwofontstrue

\ifoldfss
  \ifCUPmtlplainloaded \else
    \NewTextAlphabet{textbfit} {cmbxti10} {}
    \NewTextAlphabet{textbfss} {cmssbx10} {}
    \NewMathAlphabet{mathbfit} {cmbxti10} {} 
    \NewMathAlphabet{mathbfss} {cmssbx10} {} 
  \fi
  \ifAMStwofonts
    \ifCUPmtlplainloaded \else
      \NewSymbolFont{upmath} {eurm10}
      \NewSymbolFont{AMSa} {msam10}
      \NewMathSymbol{\upi}     {0}{upmath}{19}
      \NewMathSymbol{\umu}     {0}{upmath}{16}
      \NewMathSymbol{\upartial}{0}{upmath}{40}
      \NewMathSymbol{\leqslant}{3}{AMSa}{36}
      \NewMathSymbol{\geqslant}{3}{AMSa}{3E}

    \fi
  \fi
\fi 

\ifnfssone
  \newmathalphabet{\mathit}
  \addtoversion{normal}{\mathit}{cmr}{m}{it}
  \addtoversion{bold}{\mathit}{cmr}{bx}{it}
  \newmathalphabet{\mathbfit} 
  \addtoversion{normal}{\mathbfit}{cmr}{bx}{it}
  \addtoversion{bold}{\mathbfit}{cmr}{bx}{it}
  \newmathalphabet{\mathbfss} 
  \addtoversion{normal}{\mathbfss}{cmss}{bx}{n}
  \addtoversion{bold}{\mathbfss}{cmss}{bx}{n}
  \ifAMStwofonts
    \ifCUPmtlplainloaded \else
      \UseAMStwoboldmath
      \makeatletter
      \new@mathgroup\upmath@group
      \define@mathgroup\mv@normal\upmath@group{eur}{m}{n}
      \define@mathgroup\mv@bold\upmath@group{eur}{b}{n}
      \edef\UPM{\hexnumber\upmath@group}
      \new@mathgroup\amsa@group
      \define@mathgroup\mv@normal\amsa@group{msa}{m}{n}
      \define@mathgroup\mv@bold\amsa@group{msa}{m}{n}
      \edef\AMSa{\hexnumber\amsa@group}
      \makeatother
      \mathchardef\upi="0\UPM19
      \mathchardef\umu="0\UPM16
      \mathchardef\upartial="0\UPM40
      \mathchardef\leqslant="3\AMSa36
      \mathchardef\geqslant="3\AMSa3E
    \fi
  \fi
\fi 

\ifnfsstwo
  \DeclareMathAlphabet{\mathbfit}{OT1}{cmr}{bx}{it}
  \SetMathAlphabet\mathbfit{bold}{OT1}{cmr}{bx}{it}
  \DeclareMathAlphabet{\mathbfss}{OT1}{cmss}{bx}{n}
  \SetMathAlphabet\mathbfss{bold}{OT1}{cmss}{bx}{n}
  \ifAMStwofonts
    \ifCUPmtlplainloaded \else
      \DeclareSymbolFont{UPM}{U}{eur}{m}{n}
      \SetSymbolFont{UPM}{bold}{U}{eur}{b}{n}
      \DeclareSymbolFont{AMSa}{U}{msa}{m}{n}
      \DeclareMathSymbol{\upi}{0}{UPM}{"19}
      \DeclareMathSymbol{\umu}{0}{UPM}{"16}
      \DeclareMathSymbol{\upartial}{0}{UPM}{"40}
      \DeclareMathSymbol{\leqslant}{3}{AMSa}{"36}
      \DeclareMathSymbol{\geqslant}{3}{AMSa}{"3E}
    \fi
  \fi
\fi 

\ifCUPmtlplainloaded \else
  \ifAMStwofonts \else 
    \def\upi{\pi}
    \def\umu{\mu}
    \def\upartial{\partial}
  \fi
\fi

\title[Brown dwarfs and low-mass stars in the Pleiades and Praesepe]
{Brown dwarfs and low-mass stars in the Pleiades and Praesepe:- Membership and binarity}

\author[D. J. Pinfield et al.]
  {D. J. Pinfield,$^1$ P. D. Dobbie,$^2$ R. F. Jameson$^2$, I. A. Steele$^1$, H. R. A. Jones$^1$, \\
  \newauthor and A. C. Katsiyannis$^3$ \\
  1. Astrophysics Research Institute, Liverpool John Moores University,
  Twelve Quays House, Egerton Wharf, Birkenhead, CH41 1LD \\
  2. Dept. of Physics \& Astronomy, University of Leicester,
  University Road, Leicester, LE1 7RH \\
  3. Dept. of Pure \& Applied Physics, The Queen's University of Belfast, 
  BT7 1NN, Northern Ireland
}

\date{Accepted 2003 ?;
      Received 2003 ?;
      in original form 2002 August}

\pagerange{\pageref{firstpage}--\pageref{lastpage}}
\pubyear{2003}

\begin{document}

\maketitle

\label{firstpage}

\begin{abstract}
We present near infrared J-, H- and K-band photometry and optical spectroscopy of low-mass star 
and brown dwarf (BD) candidates in the Pleiades and Praesepe open clusters. We flag non-members 
from their position in K, I-K and J, J-K colour-magnitude diagrams (CMDs), and J-H, H-K two-colour 
diagrams. In general, the theoretical NextGen isochrones of the Lyon Group fit the K, I-K CMDs well 
for stars with I-K$\sim$1.5-3.5. However Pleiades stars with K$\simeq$10.5--13 (M$_K$$\simeq$5--7.5) 
are rather redder than the isochrones. We also identify this effect amongst $\alpha$Per sources 
from the literature, but find no evidence of it for field stars from the literature. The NextGen 
isochrones fit the J, J-K CMDs of both clusters very well in this photometric range. It is possible 
that the I-K colour of youthful stars is affected by the presence of magnetic activity. The Lyon Group's 
Dusty isochrones fit both K, I-K and K, J-K Pleiades CMDs well for I-K$\simeq$4.3-6/J-K$\simeq$1.1-1.4. 
In between these colour ranges the Pleiades cluster sequence comprises three portions. Starting at 
the bluer side, there is a gap where very few sources are found (the gap size is $\Delta$I$\sim$0.5, 
$\Delta$J$\sim$$\Delta$K$\sim$0.3), probably resulting from a sharp local drop in the magnitude-mass 
relation. Then the sequence is quite flat from I-K$\sim$3.5--4. Finally, the sequence turns over and 
drops down to join the Dusty isochrone. We also compare model atmosphere colours to the two-colour 
diagrams of the clusters. The NextGen models are seen to be $\sim$0.1 too blue in H-K, and $\sim$0.1 
too red in J-H for T$_{eff}>$4000K. However, they are in reasonable agreement with the data at 
T$_{eff}\sim$3200K. For T$_{eff}\sim$2800--3150K, the colour of Pleiades and Praesepe sources are 
significantly different, where Praesepe sources are $\sim$0.1 bluer in J-H and up to $\sim$0.1 redder 
in H-K. These differences could result from gravity sensitive molecular opacities. Cooler Praesepe 
sources then agree well with the dusty models, suggesting that dust is beginning to form in Praesepe 
sources around 2500K. However, Pleiades sources remain consistent with the NextGen models (and 
inconsistent with the dusty models) down to T$_{eff}$s of $\sim$2000K. It is possible that dust 
formation does not begin until lower T$_{eff}$s in sources with lower surface gravities (and hence 
lower atmospheric pressures). We also identify unresolved binaries in both clusters, and estimate mass 
ratios (q) for Pleiades BDs. Most of these have q$>$0.7, however, 3/18 appear to have lower q values. 
We determine the binary fraction (BF) for numerous mass ranges in each cluster, and find that it is 
generally rising towards lower masses. We find a BD BF of 50$^{+11}_{-10}$\%. We also find some 
evidence suggesting that the BF-q distribution is flat for 0.5-0.35M$_{\odot}$, in contrast to solar 
type stars.
\end{abstract}


\section{Introduction}
Low-mass star and brown dwarf (BD) populations in open clusters make excellent test-beds 
for our theoretical understanding of these objects. With a well constrained cluster age, 
and assuming uniform composition, we can effectively compare theoretical isochrones to observed 
colour magnitude diagrams (CMDs) and two-colour diagrams without the complications introduced 
by a spread in metallicity and a range in surface gravities for a given mass. Furthermore, for 
most open clusters, the depth effect is small, and the single star sequence is therefore 
quite tight on CMDs. It is thus possible to study unresolved binarity amongst cluster members.

The T$_{eff}$ range of young to intermediate age low-mass stars and BDs corresponds to M and L 
spectral types, and modeling the atmospheres of such objects is very complicated. The convection 
zone penetrates deeply into the optically thin outer layers due to H$_2$ dissociation (Burrows et 
al. 1993). Molecular opacities such as TiO and H$_2$O comprise millions of lines that must be 
accurately modeled (Allard et al. 1997). Also, dust begins to form in atmospheres at around 2800K 
(Allard et al. 2001), which depletes opacities as well as affecting the atmospheric temperature 
gradient. There are many different varieties of dust (Sharp \& Huebner 1990), and their formation 
rates and grain size distributions will depend not only on temperature, but on surface gravity 
and metallicity. Despite these complexities, much progress has been made in modeling these 
atmospheres. The Lyon Group has combined structural models (using interior structure physics) 
with model atmospheres (Hauschildt, Allard \& Baron 1999; Allard et al. 2001) to calculate 
evolutionary models for low-mass stars and BDs, both with and without atmospheric dust condensates 
(Baraffe et al. 1998, hereafter NextGen; Chabrier et al. 2000, hereafter Dusty). Calculations by 
other groups include those of Burrows et al. (1997), and D'Antona \& Mazzitelli (1994). However, 
only the Lyon group models are based on consistent non-grey evolutionary calculations relating the 
mass and age of an object to its observational properties (colour and magnitude).

The Pleiades and Praesepe are two nearby, rich open clusters with near solar composition 
(Crawford \& Perry 1976; Hambly et al. 1995b). With distances of $\sim$125pc and 160pc 
respectively, both clusters have small depth effects, and are fairly compact on the sky 
(a few degrees across). The Pleiades is $\sim$125Myrs old (Stauffer, Schultz \& Kirkpatrick 
1998), and Praesepe is $\sim$0.6-1.4Gyrs old (Hambly et al. 1995b). As a consequence, Pleiades 
low-mass stars ($le$0.3M$_{\odot}$) and BDs are not fully contracted, and will have lower 
surface gravities than their equal T$_{eff}$ counterparts in Praesepe.

In this paper, we derive membership criteria of Pleiades and Praesepe low-mass star and BD candidates 
using near infrared photometry and optical spectroscopy. We compare NextGen and Dusty theoretical 
isochrones to our observed CMDs and two-colour diagrams, and identify photometric ranges where the 
isochrones agree well with observations. We then join these isochrone sections together using additional 
empirical points, and discuss the overall form of the cluster sequences. We also use our CMDs and 
two-colour diagrams to identify unresolved binaries. We derive binary fractions over several mass 
ranges in each cluster, and estimate mass ratios for the Pleiades BD binaries. Finally, we discuss 
future work.

\section{Observations \& data reduction}

\subsection{The Pleiades \& Praesepe samples}
Our Pleiades and Praesepe samples were taken from several sources. For the Pleiades, the highest 
mass sources ((I$<$13.5) were taken from Hambly, Hawkins \& Jameson (1993; HHJ hereafter). Fainter 
sources were taken from, the  ITP survey (Zapatero-Osorio et al. 1999), the CFHT survey (Bouvier 
et al. 1998), the Burrell Schmidt survey (Pinfield et al. 2000; BPL hereafter) and the INT WFC survey 
(Dobbie et al. 2002b). The fainter surveys cover a combined area of 7.6 square degrees, and range in 
photometric depth from I=19.5-21.8 (unless otherwise stated, I refers to Cousins I). These fainter 
surveys are summarised in more detail in Jameson et al. (2002).

Our Praesepe sample was taken from four surveys. Proper motion members from the 19 square degree 
survey of Hambly et al. (1995a, hereafter HSHJ) represent our highest mass sources (I=10-17.5). 
Lower mass sources come from Pinfield et~al. (1997), Pinfield (1997) and Magazzu et al. (1998). 
The Pinfield surveys covered 1 and 6 square degrees, and reached I=21 and 19.5 respectively. The 
Magazzu survey covered 800 square arcminutes down to I=21.2, and identified 1 BD candidate which 
they called Roque~Pr~1.

\subsection{Photometry}

\begin{table*}
\caption{Spectral indices and equivalent widths of the Pleiades BD candidates.}
\begin{center}
\begin{tabular}{|l|c|c|c|c|c||c|c|c|c|c}
\hline
Source & SpT & PC3 & A & VO & I3 & I2 & H$_{\alpha}$EW & KI 7655 & KI 7699 & NaI 8183+8195 \\
&&&&&&& (\AA) & EW (\AA) & EW (\AA) & EW (\AA) \\
\hline
BPL 327& dM7.1 & 1.55 & 1.52 & 2.45 & 2.33 & 1.03 & -5$\pm$2 & -12$\pm$3 & -10$\pm$3 & -7$\pm$1 \\
BPL 45 & dM8.1 & 1.76 & 1.69 & --   & --   & --   & --       & --        & --        & -7$\pm$2 \\
\hline
\end{tabular}
\end{center}
\end{table*}
For the brighter BPL and HSHJ candidates we obtained J- H- and K-band magnitudes from 
the 2MASS 2nd incremental point-source catalogue at the NASA/IPAC InfraRed Science 
Archive (Skrutskie et al. 1995). We used a search radius of 3 arcseconds to allow for 
positional uncertainties. For the ITP, CFHT and INT WFC surveys, we obtained photometry 
from the literature for as many sources as was available (see Jameson et al. 2002). Most 
of the fainter BPL candidates had K-band measurements from Pinfield et al. (2000), and 
several Praesepe candidates had K-band measurements from Hodgkin et al. (1999). For the 
remainder of these samples we measured J-, H- and K-band photometry using UFTI on the 
United Kingdom Infrared Telescope (UKIRT) on Mauna Kea, Hawaii, from 3--5 January 2000, 
and in service time on 8 October 2001. The sky conditions were photometric. Our observing 
strategy for candidates without a K-band magnitude was to observe them first at K, and 
only go on to measure J- and H-band magnitudes if the targets had a sufficiently red I-K 
colour to be consistent with membership of the cluster. For the remaining candidates, we 
simply measured J- and H-band magnitudes, as well as K-band if the previous K uncertainty was 
$>$0.1. Each observation comprised a five-point dither pattern, with a windowed 512$\times$
512 pixel readout. Total integration times at each of J, H and K were 150s, 300s and 540s, for 
objects with I=17--18.5, I=18.5--20 and I$>$20 respectively. The raw images were de-biased, 
dark subtracted, flat-fielded, and combined into mosaics using {\sc ORACDR}. Aperture 
photometry was then extracted using {\sc GAIA}. As the UKIRT photometry was obtained with the 
``Mauna Kea Observatory'' (MKO) filter set (Simons \& Tokunaga 2002), the photometry from the 
literature was transformed into the MKO system using the transforms of Hawarden et al. (2001) 
and Carpenter (2001) (old UKIRT system and 2MASS respectively). The photometry is given in 
the appendices. Where there were multiple measurements available of an object, we selected 
those with small errors (generally trying to avoid measurements with errors larger than 
$\sim$0.1), and took an average.

Our optical photometry was taken from the literature. All I-band photometry was transformed 
into the Cousins system. Photographic I$_N$ magnitudes from HSHJ were transformed using Bessell 
(1986). To transform the I$_{KP}$ magnitudes of Pinfield (1997) and Pinfield et al. (2000) we 
derived the following relation; I$_C$-I$_{KP}$=0.049(I$_{KP}$-K)+0.069 (for $2.3<I_C-K<5.8$). 
This relation was obtained using synthetic photometry derived from spectra, using the method 
and data from Dobbie et al. (2002a).

\subsection{Spectroscopy}

We obtained spectra of 2 BD candidates using ISIS on the 4.2~m William Herschel Telescope 
(WHT) during 29 Jan 2001. The seeing was $\sim$1.2 arcseconds and the weather was clear. Our 
instrumental setup comprised the TEK4 CCD and the R158R grating on the ISIS red arm, with the 
mirror in place and no order blocking filters, so as to minimize flux losses (we expect second 
order contamination for these red objects to be negligible). We used a 1 arcsecond slit, and 
integrated for 1800s. This gave us a wavelength coverage of 6213-9193\AA\, with a dispersion 
of 2.91 \AA\ pixel$^{-1}$.

We used IRAF to reduce our spectra. The CCD frames were bias-subtracted and flat-fielded 
using standard routines, and the spectra extracted and sky-subtracted using the {\it apall} 
routine. We calibrated the wavelength scale using the CuAr arc lamp exposures, and the flux 
using observations of flux standard stars (available in the IRAF environment) on the same night, 
and with the same instrumental setup.

\section{Results \& discussion}

\subsection{Pleiades spectra}

\begin{figure}
 \epsfig{file=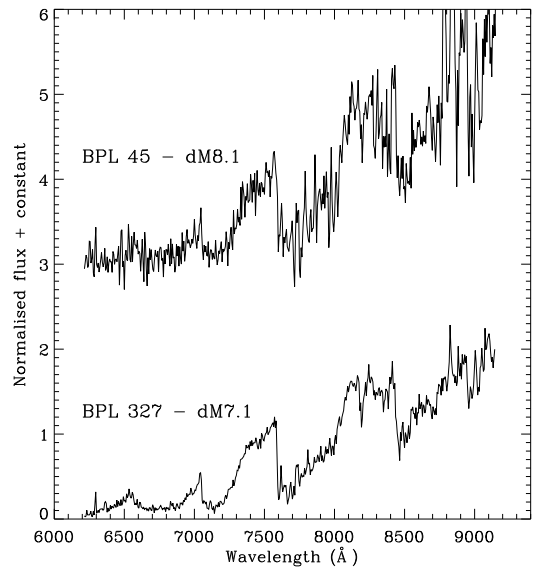,width=8.0cm}
 \caption{The 2 Pleiades BD candidate spectra (normalised to unity at 7500\AA). The noisier 
spectra of BPL 45 has been binned by 2 pixels in this plot, and offset for clarity.}
\end{figure}

Figure 1 shows the spectra of the two Pleiades BD candidates BPL 327 and BPL 45, where they 
have been normalised to unity at 7500\AA. Table 1 shows the spectroscopic information extracted 
from these data. Our adopted spectral types were derived using the Pseudo-continuum ratio PC3 
from Mart\'{\i}n, Rebolo \& Zapatero-Osorio (1996), and have uncertainties of $\pm$0.5. For 
BPL 327 the VO index of Kirkpatrick, Henry \& Simons (1995) agrees with the PC3 index. Also, 
the A-index of Kirkpatrick, Henry \& McCarthy (1991) implies a slightly earlier spectral type 
for both sources (dM6.5 and dM7.5 for BPL 327 and BPL 45 respectively), which is what we expect 
for young sources that are not fully contracted. We also note that the BPL 327 spectra shows 
that the I2 and I3 indices of Mart\'{\i}n \& Kun (1996) (which measure the CaH and the TiO 
absorption bands respectively) continue to increase out to $\sim$dM7, before turning over and 
decreasing to lower values by dM8. By comparing the I magnitudes and derived spectral types with 
the spectral type-magnitude diagram of Mart\'{\i}n et al. (2000), we found that both BPL 327 and 
BPL 45 lie on the Pleiades sequence, and appear to be single BDs.

\subsection{The IK CMDs}

Figures 2 and 3 show the K, I-K CMDs (IK CMDs) for the Pleiades and Praesepe respectively. 
Small filled circles are higher mass candidate members with 2MASS photometry, and proper 
motion candidates from HSHJ in the Pleiades and Praesepe respectively. We have several 
additional fainter Praesepe candidates with 2MASS photometry which are not plotted, since 
their errors are sizable (see Table A.4). Larger filled symbols are lower mass candidates 
(candidate BDs in the Pleiades), where circles have J- and H-band photometry (triangles do 
not). Open circles (small and large) look like background sources in these plots. Open 
diamonds and open squares are likely old-disk non-members from the K, J-K CMDs (JK CMDs) and 
the J-H, H-K 2-colour diagrams (JHK diagrams) respectively (see Sections 3.4 and 3.5). Circled 
objects look like unresolved binaries in these plots (see below). Objects over-plotted with an 
additional square or triangle are likely unresolved binaries from the JK CMDs and JHK diagrams. 
Typical photometric error bars are shown in the bottom left. The location of several magnitude 
bins are indicated with dotted lines (see Section 3.5). Solar metallicity NextGen and Dusty 
isochrones are over-plotted as dashed lines, with mass points shown in units of 0.001M$_{\odot}$. 
We also show higher mass ($>$0.7M$_{\odot}$) isochrones derived with a larger mixing length 
parameter ($l$=1.9$H_p$), since this parameter makes a significant difference at these higher 
masses (Chabrier \& Baraffe 1997). In the Pleiades, we assume an age of 125Myrs, A$_I$=0.06, 
A$_K$=0.05, and use the re-computed Hipparcos distance modulus of (m-M)$_0$=5.57 (Makarov 2002), 
which was derived in a way that reduces the propagation of along-scan attitude errors (a source 
of discrepancy with the previous Hipparcos Pleiades distance (eg. van Leeuwen 1999). In Praesepe, 
we show both 0.5 and 1Gyr isochrones, and assume (m-M)$_0$=6.0 with zero extinction. This distance 
modulus was obtained from a best-fit of the NextGen isochrones to the good Praesepe candidates 
with UKIRT photometry from I-K=2.5--3.5. In this colour range the 0.5 and 1Gyr isochrones lie 
on top of each other. This distance agrees well with other values found from main sequence fitting 
(Lynga 1987, Nissen 1988), but is slightly lower than the Hipparcos value of 6.24 (Mermilliod et al. 
1997). However, it is possible that the previously mentioned attitude errors could affect the 
Hipparcos distance of Praesepe. The over-plotted solid lines represent our selected cluster sequences 
(constructed both from isochrone data and empirical data), with their equal mass binary sequences 
shown as dotted lines (see subsequent discussion).

\begin{figure}
 \epsfig{file=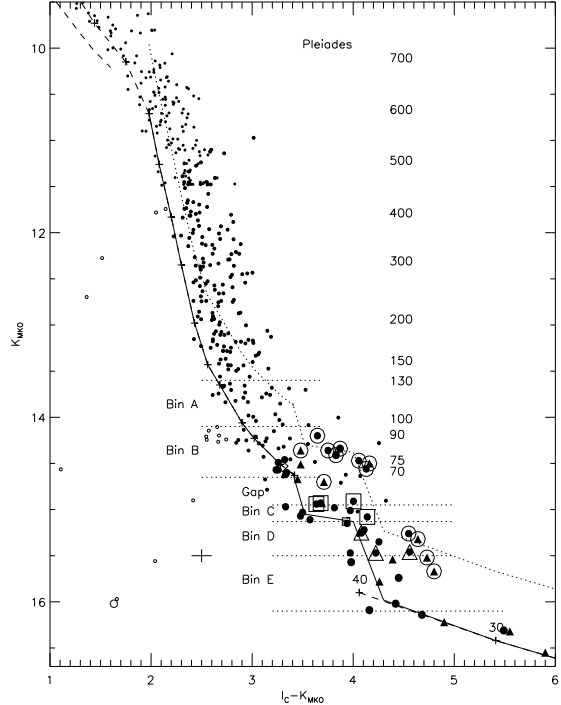,width=8.0cm}
 \caption{The K, I-K CMD of the Pleiades. Small filled circles have 2MASS photometry. Larger 
filled symbols have UKIRT photometry; circles have JHK, triangles have just K. Open circles are 
non members. Circled objects are IK binaries. Objects over-plotted with an additional open square 
or triangle are JK or JHK binaries respectively (see text). Dashed lines are the NextGen and Dusty 
models. Solid and dotted lines are the cluster single and binary star sequences respectively.}
\end{figure}

\begin{figure}
 \epsfig{file=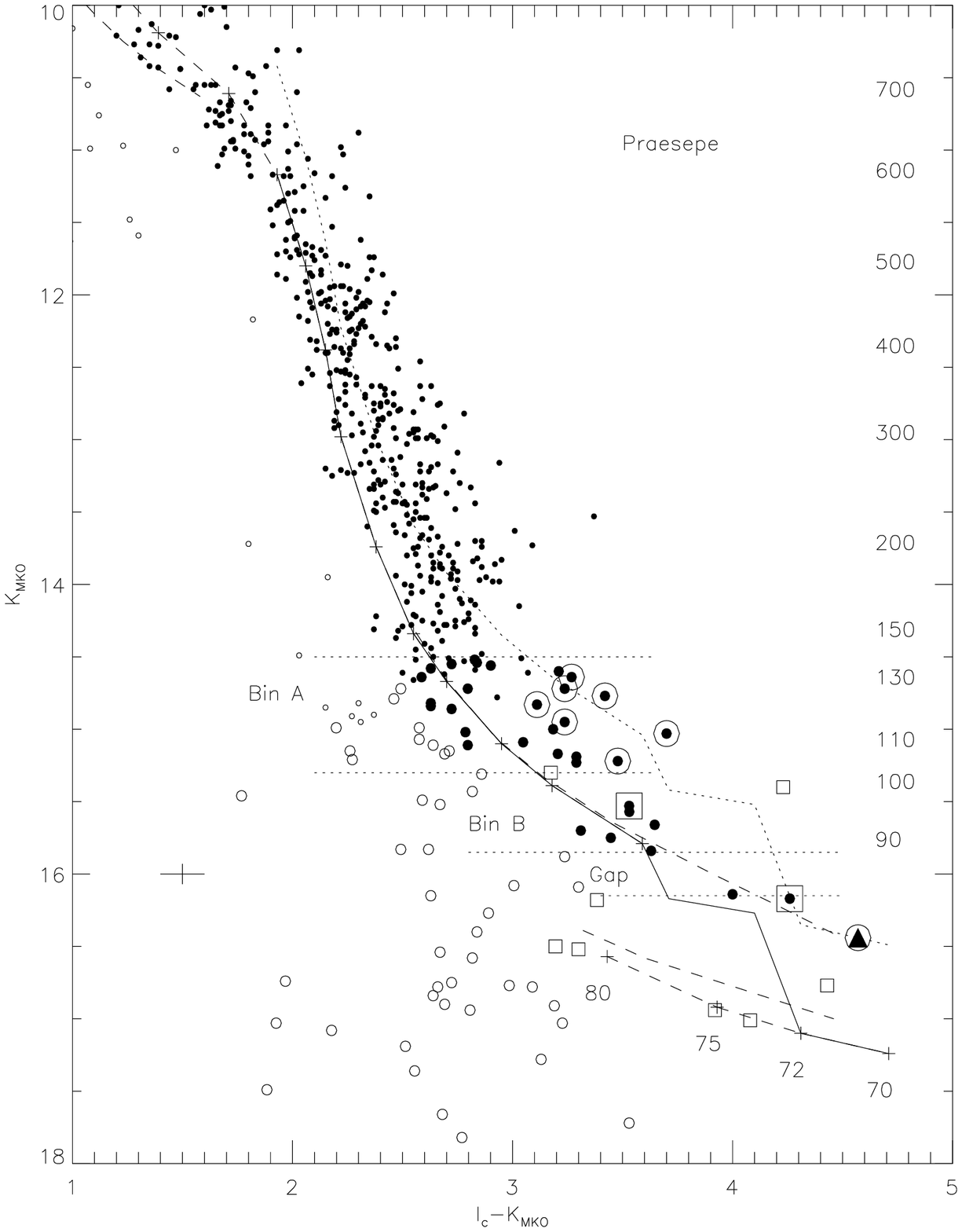,width=8.0cm}
 \caption{The K, I-K CMD of Praesepe. Symbols are the same as in Figure 2.}
\end{figure}

\begin{figure}
 \epsfig{file=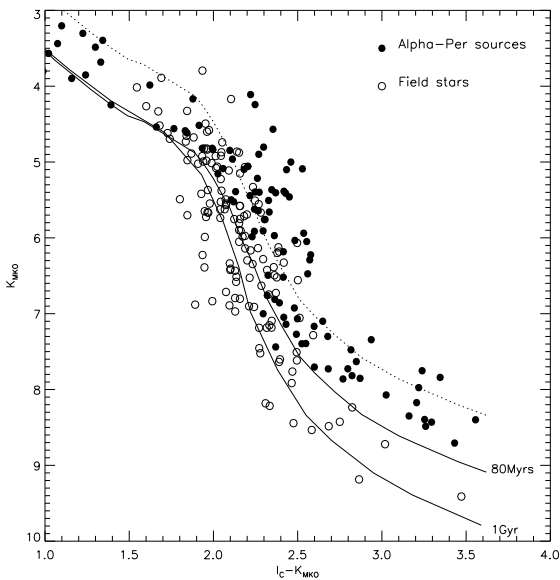,width=8.0cm}
 \caption{The K, I-K CMD of Alpha-Per (filled circles) and field stars (open circles). 
Sources are taken from Prosser (1992), Barrado y Navascu\'es (2002), and Reid \& Cruz (2002).}
\end{figure}

We will first discuss the brighter sources with 2MASS K-band photometry. 
For the range K=10.5--14 in the Pleiades, and K=11--14.5 in Praesepe there is a known kink 
in the main sequence, where it becomes steeper. This feature is seen over a wide range of wavelengths 
(Reid \& Cruz 2002), and is thought to result from H$_2$ formation in the photosphere, lowering the 
adiabatic gradient and causing convection (Kroupa Tout \& Gilmore 1990). The resulting lower 
temperature gradient then leads to increased T$_{eff}$s, with the consequent reduction in the I-K 
colour, and turn over on the CMD. The kink is clearly seen in both cluster CMDs. For Praesepe, the 
model isochrone follows the data very well (to within the photometric errors). However, for the 
Pleiades the data appear to be somewhat redder than the isochrones from K=10.5--13 (M$_K$$\sim$5--7.5). 
In order to determine if this effect is an artifact of the different photometric filter systems used, 
we have examined our photometry more closely. Photographic magnitudes (used in HHJ) and KittPeak 
magnitudes (used in Pinfield etal 2000) were both available for the Pleiades sources in this range, 
and by transforming both onto the Cousins system we looked for any systematic differences that could 
account for such an effect. However, no such differences were found, and the Pleiades sources looked 
redder than the isochrones whichever I-band data was used. The Pleiades and Praesepe sources in this 
range all have 2MASS K-band photometry (converted onto the MKO system), and so there should be no 
systematic difference in this band. It thus seems unlikely that this effect is due to photometric 
inconsistencies.

Assuming this effect is genuine, we should expect to see it amongst other similar populations. The 
open cluster $\alpha$Per has a similar age to the Pleiades ($\simeq$80Myrs), and Figure 4 shows an 
IK CMD of likely $\alpha$Per members (filled circles) from Prosser etal (1992) and Barrado y Navascu\'es 
(2002). These have been corrected for distance and reddening (m-M=6.23, A$_I$=0.17, A$_K$=0.04; Stauffer 
etal 1999) for easy comparison with the field stars that are also plotted (open circles; Reid \& 
Cruz, 2002). NextGen isochrones of 80Myrs and 1Gyr are indicated by solid lines, and a dotted line 
indicates the expected spread due to binarity for $\alpha$Per. It is clear that (despite some scatter) 
the 1Gyr model isochrones represents the field star data very well. However, the majority of the 
$\alpha$Per sources from M$_K\sim$5-7 are redder than both the single and the binary 80Myr 
isochrones, and the same effect is therefore evident in $\alpha$Per aswell as the Pleiades.

So what could be causing this reddening? Variability could make a source appear redder since the I- 
and K-band measurements were not made simultaneously. However, sources could equally be flaring when 
the I- or K-band measurements were made, and the overall effect should be to simply broaden the sequence 
slightly, not shift it to the red. The magnetic activity of such youthful sources may, however, provide 
the explanation for the reddening. Hawley, Tourtellot \& Reid (1999) found that dMe stars in NGC 2516 
($\sim$Pleiades age) lay redward of the fiducual main sequence in an $M_V$ verses V-I CMD. They also 
found that some molecular bands respond strongly to the presence of a chromosphere, and concluded that 
the V-I colour was affected by the presence of activity. The I-K colour and the V-I colour could be 
reddened by the same process. It may seem surprising that photospheric bands (and the photometric 
colours they influence) should be affected by the presence of a chromosphere. However, modeling the 
correct form of the magnetic field that threads the atmospheres of these stars may allow us to account 
for these effects.

Below the kink the NextGen isochrone fits both cluster CMDs well out to I-K$\sim$3.5, showing 
the expected decreasing T$_{eff}$s as degeneracy sets in. For Praesepe, we defined our cluster 
sequence out to I-K$\simeq$3.5 with the NextGen model isochrone, since it fits the data very well 
in this range. We chose to do the same for the Pleiades despite the sources being slightly redder 
than the isochrones for K=10.5-13, since we do not expect the reddening to significantly effect 
the mass-magnitude model predictions.

At redder colours we confine our discussion to the Pleiades, since we only have a few good Praesepe 
candidates in this range. It can be seen that there is a fairly obvious gap in the Pleiades sequence 
from K$\simeq$14.65--14.95 (at I-K$\simeq$3.5) where the number of sources is severely reduced. The 
spectral type corresponding to the gap is M7 (see Figure 4 of Mart\'{\i}n et al. 2000), and we refer 
to this feature as the ``M-dwarf gap''. It is discussed in more detail in Dobbie et al. (2002c). 
However, to summarise, we believe that it results from a sharp local drop in the magnitude-mass 
relation, possibly caused by the formation of dust in the atmospheres of objects in this T$_{eff}$ 
regime. We measured the size of the gap at 0.3 in K which compares to $\sim$0.5 in I (Dobbie et al. 
2002c).

Below the Pleiades M-dwarf gap (K$\simeq$14.95--15.3) the single source candidates follow a fairly flat 
path in the CMD out to I-K$\sim$4.1. There is then a turn down, where the candidates drop onto the Dusty 
isochrone, joining it somewhere from I-K$\sim$4.1--4.5. We estimated the path followed by the flat 
part of the sequence by averaging the colours and K-band magnitudes of two groups of sources; the 3 
bluest single sources and the 3 reddest single sources from K=14.95--15.3 (below which the turn down 
is obvious). And we estimated the colour where the sequence joins the Dusty isochrone as I-K$\sim$4.3. 
The Dusty isochrone then agrees well with the 5 faintest Pleiades candidates out to I-K$\sim$6, so 
in this range we used the Dusty isochrones for our sequence.

For Praesepe, we have extended the cluster sequence beyond the NextGen isochrones by assuming it has 
the same form as the Pleiades sequence; the same M-dwarf gap size, and the same flat portion 
that turns over and falls onto the Dusty isochrone by I-K$\sim$4.3. This may not be entirely accurate, 
but should act as a reasonable approximation.

We have flagged non-members in the Pleiades and Praesepe using our cluster sequences as a guide, and 
allowing for photometric uncertainties as well as the depth effect of the clusters ($\pm$0.2$^{\rm mag}$ 
for the Pleiades, and $\pm$0.15$^{\rm mag}$ for Praesepe; Pinfield, Jameson \& Hodgkin 1998, Holland 
et al. 2000). We have also flagged likely unresolved binary members with UKIRT photometry, if they lie 
significantly ($>$0.25$^{\rm mag}$) above the single star sequence. We duly identified 13 Pleiades 
binaries, and 8 Praesepe binaries in this way. We refer to them as IK binaries, and they are overplotted 
with an open circle in the figures. We note that there is a degeneracy between mass and binarity where 
the IK sequence is dropping down onto the Dusty isochrone (I-K=3.9-4.5), and it is not possible to 
identify all binaries in this region from the IK CMD alone.

\subsection{Model J-K colours}

\begin{figure}
 \epsfig{file=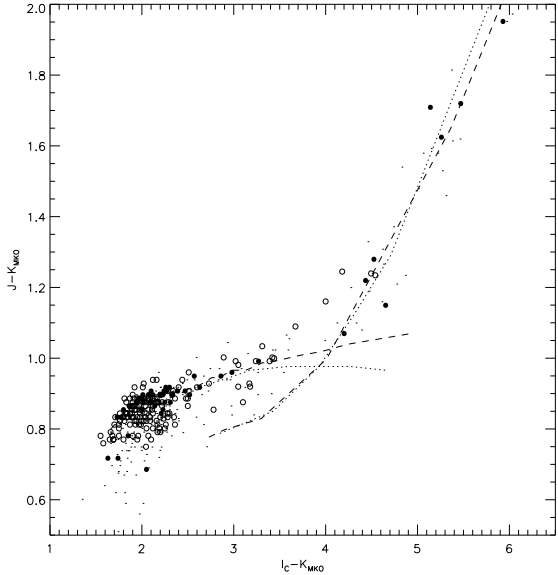,width=8.0cm}
 \caption{The J-K, I-K two-colour diagram of field star sources. Filled circles are young disk, and 
open circles are old disk. Points have no age information available. 125Myr and 1Gyr NextGen and Dusty 
isochrones are shown as dashed and dotted lines respectively.}
\end{figure}

In the previous section we found that the NextGen isochrones fit the IK CMDs very well for 
I-K$\sim$1--3.5, except for Pleiades sources from K=10.5--13, which are slightly redder than 
the isochrones. We also found that Dusty isochrones fit the Pleiades IK CMD well for I-K=4.3--6. 
In this section we test the J-K model predictions for these ranges, for a sample of field stars 
with kinematics and well measured I-K and J-K colours.

Figure 5 shows J-K plotted against I-K for sources taken from Leggett (1992), Leggett, Allard \& 
Hauschildt (1998) and Dahn et al. (2002). All sources classified as young or old disk (from their 
kinematics) in the Leggett papers are shown as filled or open circles respectively. In addition, 
we have assigned 4 L dwarfs as young ($<$0.5Gyrs), based on lithium detections (Mart\'{\i}n et al. 
1999). These are also shown as filled circles. The uncertainties are generally $\pm$0.05 in colour. 
We have also overplotted the NextGen and Dusty model isochrones for 120Myrs (dashed lines) and 
1Gyrs (dotted lines).

Comparison of the two populations shows that for bluer sources (I-K$<$3.5, J-K$<$1), the old disk stars 
have a bigger spread in J-K and on average are slightly bluer in J-K than their young disk counterparts. 
The two NextGen isochrones essentially lie on top of each other in this range, and fit the data very 
well. For redder sources (I-K=4.3--6, J-K=1.1--2), the Dusty isochrones lie on top of each other, and 
are in excellent agreement with the young disk data. In between these two ranges the isochrones are all 
too blue in J-K. The NextGen models are too blue because they do not account for dust, and the Dusty 
models are too blue because they use a less sophisticated treatment of H$_2$O and TiO molecular opacities.
This is all consistent with the gradual change-over from NextGen to Dusty isochrones that was seen 
in the IK CMDs, and suggests that these models should provide accurate NIR isochrones in the blue and 
red photometric ranges.

\subsection{The JK CMDs}

\begin{figure}
 \epsfig{file=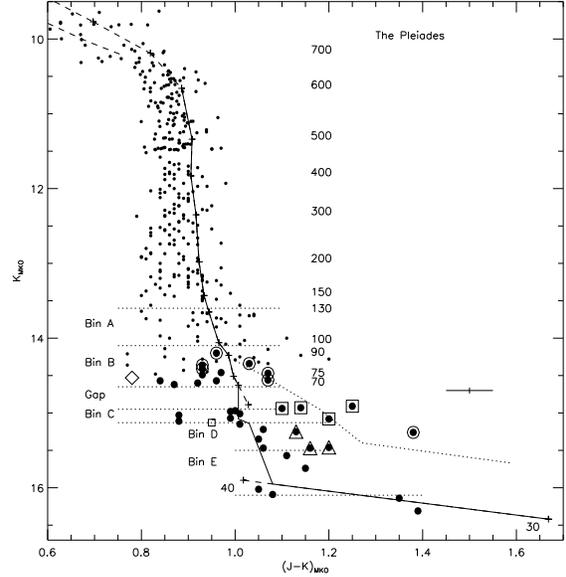,width=8.0cm}
 \caption{The K, J-K CMD of the Pleiades. Symbols are the same as in Figure 2.}
\end{figure}

\begin{figure}
 \epsfig{file=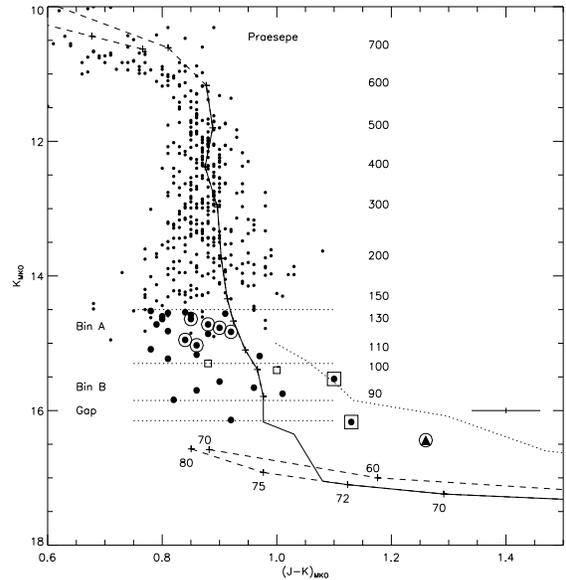,width=8.0cm}
 \caption{The K, J-K CMD of Praesepe. Symbols are the same as in Figure 2.}
\end{figure}

Figures 6 and 7 show the JK CMDs for the Pleiades and Praesepe respectively. Only good candidates 
from the IK CMDs are shown. The symbols mean the same as in Figure 2. The location of several 
magnitude bins are indicated with dotted lines (see Section 3.5). Isochrones are overplotted as 
dashed lines, with mass points shown in units of 0.001M$_{\odot}$. Typical photometric errors for 
UKIRT photometry are indicated.

We constructed cluster sequences for the JK CMDs based on our analysis in Sections 3.2 and 3.3. We 
used NextGen isochrones for I-K$<$3.5 (mass$<$0.07 and 0.09M$_{\odot}$ for the Pleiades and Praesepe 
respectively), and Dusty isochrones for J-K=1.1--2 (I-K=4.3--6). In between the NextGen and Dusty 
isochrones we added two additional sequence points, based on our Pleiades sources; one to bridge the 
M dwarf gap, and the other to represent the turn down point, where the sequences begins to fall onto 
the Dusty isochrone (see Section 3.2). An average of the J-K colours of the Pleiades sources above 
and below the gap shows that there is essentially no significant J-K colour change across the gap 
($\Delta$J-K=0.01). Averaging of the 3 single Pleiades sources in the turn down region gave a turn 
down colour of J-K=1.03. The flat region seen in the IK CMDs is thus not very apparent in the JK 
CMDs. For Praesepe we assumed that the JK colour and magnitude changes are the same as in the 
Pleiades (as we did previously for the IK CMDs). The cluster sequences we constructed are shown 
as solid lines in the JK CMDs, and a dotted line indicates the spread we expect in the sequence 
from unresolved binarity.

The NetGen part of the sequence fits the observations very well in both clusters, and there is 
certainly no evidence for the Pleiades sources being redder than the isochrones from K=10.5-13, as 
was seen in the IK CMDs. The J-K colour is clearly not affected in the way that the I-K colour is. 
Below the M-dwarf gap, it can be seen that there is a spread in the Pleiades sources, where 
several are significantly redder than expected for single objects. A spread is also seen for a small 
number of Praesepe sources. Such a spread could result from several possibilities. Transient surface 
features such as spots could be responsible. The 2 most obvious red sources (Teide 1 and Calar 3; 
BPL 137 and 235) have been tested for variability. Bailer-Jones \& Mundt (2001) found no evidence 
of I-band variability in Teide 1 over a 100 hour time scale, and only marginal evidence (just 
above the noise) for sporadic variability ($\Delta$I$<$0.03) in Calar 3 over a 29 hour time 
scale. Our photometry of Teide 1 comprises 3 well separated ($>$a year) J- and H-band measurements 
which are consistent to within their errors. For Calar 3 we have 1 K-band measurement which is 
consistent with the measurement presented by Mart\'{\i}n et al. (2000), and 2 H-band measurements 
which agree to within their errors. We also have 3 J-band measurements which show marginal evidence 
(at the 2$\sigma$ level) of variability. The level of this variation is not in itself sufficient 
to account for the redness of Calar 3. Photometric errors will also cause some spread in the 
sources. However, the general consistency (at the 1-$\sigma$ level) between our multiple 
measurements makes it unlikely that these alone are responsible. The third option, and the one we 
chose to consider in more detail, is that additional unresolved binaries (not identified in the IK 
CMD) are present. A substantially redder unresolved companion could redden a source sufficiently to 
account for the spread. These binaries would either be background sources, or cluster binaries that 
come from the turn down region of the IK CMD, where binarity is not always apparent. We have thus 
flagged 4 additional candidate unresolved binaries in the Pleiades, and 2 in Praesepe. We refer to 
these as JK binaries, and they are overplotted with an additional open square in the figures. We note 
that the bluest of the Pleiades JK binaries is a known background binary (with out lithium; Mart\'{\i}n 
et al. 2000), resolved by the HST (Mart\'{\i}n et al. 1998). The Pleiades binaries will be discussed 
in more detail in Section 3.7. Of the 2 Praesepe JK binaries, the bluer one (RIZ~Pr 23) is located 
just above where we expect the missing M dwarf gap to be, and could be a binary from below the gap. 
The redder one (RIZ~Pr 23) comes from the turn down region of the IK CMD.

We have flagged non-members in the Pleiades and Praesepe using our cluster sequences as a guide. 
Objects that look too blue are presumably background stars from an old-disk population (see Section 
3.3), and we flagged them as such if their J-K colours were at least 2-$\sigma$ blueward of the 
cluster sequence.

\subsection{The J-H, H-K 2-colour diagrams}

Figures 8 and 9 show the J-H, H-K 2-colour diagrams (JHK diagrams) for the Pleiades 
and Praesepe respectively. Only good candidates from the IK CMDs are shown. The source 
symbols mean the same as in Figure 2. The positions of 4 field L-dwarfs (from Dahn et al. 
2002) are also shown as asterisks. The photometric error bar shown in the lower right 
of the plots are typical of the photometry of the BPL sources. The fainter sources have 
slightly larger errors. Solar metallicity model atmosphere predictions (used for the NextGen 
and Dusty models) are overplotted as dashed and solid lines respectively, with labels indicating 
surface gravities. For the Pleiades we plot NextGen models for T$_{eff}>$2000 and 3000K, and Dusty 
models for T$_{eff}<$2000K. For Praesepe, we plot NextGen models for T$_{eff}>$2500 and 3000K, 
and Dusty models for T$_{eff}<$2500K.

For the highest mass stars, it can be seen that the NextGen models are $\sim$0.1 too blue in 
H-K up to the 2-colour turn-over, where convection begins ($\sim$4000K). The data turns over at 
H-K=0.27, J-H=0.57, and $\log{g}$ for such objects should be $\sim$4.5. The theoretical J-H colour 
of the turn over is therefore $\sim$0.1 too red. These discrepancies are thought to be caused by 
inaccuracies in the modeling of water vapour opacities in the H-band (with its overall low opacity) 
as well as an underestimation of the mixing length parameter. After the turn-over, the 2MASS photometry 
follows a short path in the JHK diagram, reaching J-H=0.52, H-K=0.38 by $\sim$3200K, in reasonable 
agreement with the $\log{g}$=5 NextGen model. In order to look for any non-evolutionary differences 
between the colours of Pleiades and Praesepe stellar members, we selected Pleiades members from 
K=11.5--12, and Praesepe members from K=12--12.5. These two sets of sources should be fully contracted 
(according to the models), and should represent the T$_{eff}\sim$3500--3600K range (since we have 
corrected for distance). We see no significant difference in the colours, with J-H=0.53$\pm$0.03 and 
0.54$\pm$0.03, and H-K=0.35$\pm$0.03 and 0.33$\pm$0.04 for the Pleiades and Praesepe respectively. 
Praesepe is known to be slightly metal rich compared to the Pleiades, with [Fe/H]=0.04$\pm$0.04--0.14 
(Friel \& Boesgaard 1992; Reglero \& Fabregat 1991). However, this does not appear to significantly 
affect the stellar NIR colours, consistent with the model predictions in this T$_{eff}$ range.

\begin{figure}
  \epsfig{file=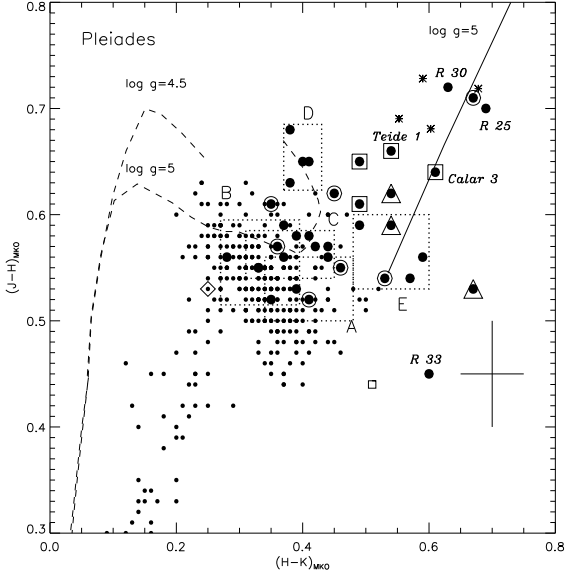,width=8.0cm}
  \caption{The J-H, H-K 2--colour diagram of the Pleiades. Symbols are the same as in Figure 2. 
  Individual sources are labeled in italic ({\it R}=Roque).}
\end{figure}

\begin{figure}
  \epsfig{file=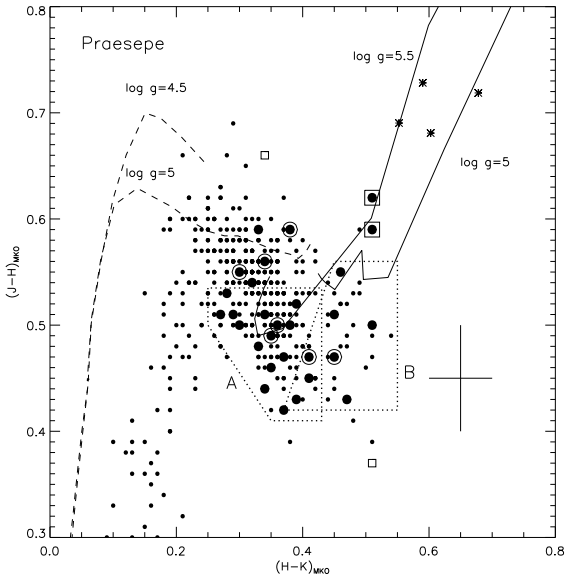,width=8.0cm}
  \caption{The J-H, H-K 2-colour diagram of Praesepe. Symbols are the same as in Figure 2.}
\end{figure}

In order to show what is happening at cooler T$_{eff}$s, we grouped the sources 
into a number of K magnitude bins. Above the gap we selected Pleiades bins of K=13.6--14.1 
and K=14.1--14.65, and Praesepe bins of K=14.5--15.3 and K=15.3--15.85. We labeled these bins 
A and B respectively, and they should correspond to approximately the same T$_{eff}$ ranges 
in each cluster ($\sim$3150--3000K for A, and 3000--2800K for B). Below the gap, we considered 
3 additional Pleiades bins. The K=14.9--15.13 range (bin C) contains objects on the flat section 
of the IK CMD (see Section 3.2). The K=15.13--15.5 range (bin D) contains objects at the top 
of the turn down, where the sequence begins to drop towards the Dusty isochrone in the IK CMD. 
Finally, the K=15.5--16.1 range (bin E) contains objects from near the base of the turn down region, 
where the IK sequence is joining onto the Dusty isochrone. These bins are indicated on the IK 
and JK CMDs with dotted lines, and in the JHK diagrams by boxes drawn with dotted lines. In 
the Pleiades, bin A contains 2MASS sources, and we defined the box dimensions using the standard 
deviation of the average bin colours. For the remaining bins we defined the box dimensions by 
attempting to include all non-binary member sources within the box dimensions.

When we compare the colour--colour location of bins A and B in each of the clusters, it can be 
seen that they differ. In the Pleiades, bins A and B have similar colours, and lie very close to 
the NextGen model predictions for their T$_{eff}$ range. However, in Praesepe these bins have J-H 
colours that are $\sim$0.1 bluer than this, and bin B has an H-K colour $\sim$0.1 redder. The surface 
gravities expected for these bins are $\log{g}$=5 for the Pleiades, and $\log{g}$=5.15--5.25 for 
Praesepe, and these differing surface gravities may provide the explanation for the differing colours. 
We do not expect any atmospheric dust formation for objects in these bins since their T$_{eff}$s are 
too high (see Tsuji 2002 figure 7, and Jones \& Tsuji 1997), so gravity sensitive dust formation cannot 
provide the explanation. However, molecular opacities (such as H$_2$O) could be more sensitive to gravity 
(or pressure) than the NextGen models predict and future generations of models with updated EOS and 
opacity data may account for these colour changes.

Moving to fainter cooler objects in Praesepe, it can be seen that the faint binary candidate below 
bin B lies on the dusty isochrone, consistent with dust formation beginning in Praesepe at around 
T$_{eff}\sim$2500K. However, in the Pleiades we are faced with a rather different story. Moving across 
the gap from bin B to C produces no significant change in colour, and between bins C and D there is 
then an increase of $\sim$0.1 in J-H. These changes are all consistent with the NextGen isochrone. 
Pleiades sources eventually move onto the dusty isochrone between bins D and E, and the 2 faintest 
Pleiades sources (Roque 25 and 30) both lie on the Dusty isochrones. A possible explanation for these 
near infrared colours is that dust does not begin to form in Pleiades sources until T$_{eff}\sim$2000K 
(between bins D and E). This might occur because in lower gravity atmospheres the lower pressures could 
suppress dust formation until lower T$_{eff}$s. However, one would also expect that lower gravity 
atmospheres would be more extended, and have lower outer temperatures, which would have the opposite 
effect.

Assuming that the differences in colour between Pleiades and Praesepe sources result from differing 
surface gravities, then one might expect even greater differences for sources found in star forming 
regions. This would have implications for the derreddening process that is commonly used for such 
sources (eg. Lucas \& Roche 2000).

We have flagged likely non-members using the location of bins A to E and the Dusty model atmosphere 
predictions (for the faintest sources) as a guide. Old disk sources appeared too blue in J-H, and 
several non-stellar sources were also identified amongst our data by their extremely red colours. 
One object in Figure 7 (Roque 33) appears to be too blue in J-H. This object is a proper motion member 
however (see Table A.1), and we note that its photometric errors are $\sim$0.2 in colour making 
them not inconsistent with membership. We also used the Pleiades JHK diagram to identify some additional 
unresolved binaries. In the turn down region of both the IK and the JK CMDs (where the sequence is 
dropping steeply down towards the Dusty isochrones), there remains some degeneracy between mass and 
binarity. However, we were able to break this degeneracy using our JHK diagram. A single source with 
K=15.13-15.5 should lie in bin D. However, 3 sources in this magnitude range lie in (or very near to) 
bin E. These are presumably near equal mass binaries, where both components of each binary come from 
bin E (we refer to them as JHK binaries). These binaries are overplotted with an additional open 
triangle in the figures.

\subsection{Summary of membership criteria}
Using the IK and JK CMDs, as well as the JHK diagrams, we have assigned photometric membership 
criterea to our Pleiades and Praesepe samples (see Appendices). In the Pleiades, we flagged 5 of 
the 30 BPL BD candidates as non-members. The remaining 25 BPL candidates, and the 9 additional 
Pleiades candidates from the literature look like cluster BDs. Examination of Table A1 shows that 
our criterea for these sources are consistent with existing spectroscopic and astrometric measurements, 
where they exist. It is particularly encouraging that we were able to flag CFHT-PL-18 (BPL 283) as a 
background binary from its photometry alone. This source was a suspected BD binary, and was only 
confirmed as a background source when Mart\'{\i}n et al. (2000) failed to detect lithium in its spectrum. 
BPL 45 has been flagged as the same type of object based on its measured J- and K-band photometry alone. 
Of the Pleiades low-mass star candidates, we flag 14 non-members out of the 194 candidates. This represents 
only 7\% spurious sources in the sample. In Praesepe, we identify higher levels of contamination for our 
faintest sources. We flag 20 non-members out of the 26 RIZ-Pr candidates, and 37 non-members out of the 
89 IZ-Pr candidates. However, for the brighter HSHJ sources we only flag 21 non-members out of 458. The 
position of Roque-Pr~1 in the IK and JK CMDs is consistent with membership. However, if it is a member, 
it must be an unresolved BD binary.

\subsection{BD binary mass-ratios}

\begin{figure*}
  \epsfig{file=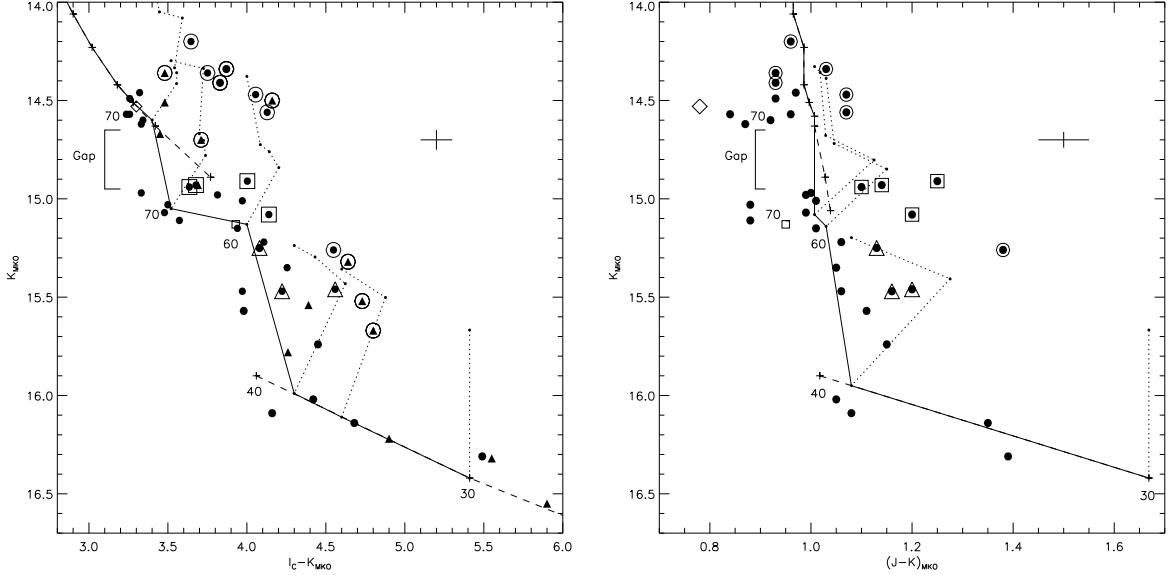,width=16.0cm}
  \caption{K, I-K and K, J-K CMDs of Pleiades BDs.}
\end{figure*}

\begin{table*}
\caption{Pleiades BD binary mass-ratios (q). The q$_1$ ranges were estimated assuming 
a depth effect of $\pm$0.1$^{mag}$, and masses from Figure 9. q$_2$ ranges were 
estimated assuming a depth effect of $\pm$0.2$^{mag}$, and accounting for the fact 
that the Dusty isochrone masses could be underestimated by $\sim$0.01M$_{\odot}$.}
\begin{tabular}{|l|l|l|l|l|l|}
\hline
Binaries & Other names & Binary type & q$_1$ & q$_2$ & Notes \\
\hline
BPL~45       &                 & JK  & & & Probable BG source \\
BPL~66       & Roque~4         & IK  & & $\sim$1 & Near front of cluster?/Triple BD? \\
BPL~79       & Roque~13        & IK  & 0.7-1 & 0.65-1 & \\
BPL~108      & Roque~14        & IK  & 0.8-1 & 0.75-1 & \\
BPL~137      & Teide~1         & JK  & 0.55-0.65 & 0.5-0.9 & \\
BPL~142      & Roque~17        & IK  & 0.7-1 & 0.65-1 & \\
BPL~235      & Calar~3         & JK  & $\sim$0.5 & 0.5-0.7 & \\
BPL~249      &                 & JHK & $\sim$1 & & \\
BPL~283      & CFHT-PL-18      & JK  & & & Known BG source \\
BPL~294      & CFHT-PL-12      & IK  & 0.6-0.9 & 0.4-1 & Primary probably above gap \\
BPL~303      & CFHT-PL-25      & JHK & $<$0.75 & $<$1 & \\
PPL~1        & Roque~15        & IK  & 0.8-1 & 0.75-1 & \\
PPL~15       & IPMBD~23, NPL35 & IK  & 0.7-1 & 0.65-1 & Known binary with q=0.85 \\
CFHT-PL-16   &                 & IK  & 0.8-1 & 0.75-1 & \\
CFHT-PL-23   &                 & JHK & $\sim$1 & & \\
INT-PL-IZ-20 &                 & IK  & & $<$0.8 & \\
INT-PL-IZ-25 &                 & IK  & $\sim$0.8 & & \\
INT-PL-IZ-42 &                 & IK  & 0.4-0.6 or $\sim$1 & 0-0.7 or $\sim$1 & Primary above or below gap? \\
INT-PL-IZ-43 &                 & IK  & 0.4-0.65 & 0-0.8 & \\
INT-PL-IZ-76 &                 & IK  & 0.9-1 & 0.8-1 & \\
\hline
\end{tabular}
\end{table*}

Figure 10 shows two regions of the Pleiades IK and JK CMDs, focusing on the candidate BDs. 
The symbols mean the same as in Figure 2. NextGen and Dusty isochrones are shown with dashed 
lines, and our cluster sequences (see Sections 3.2 and 3.4) are shown as solid lines. We 
also show the effects of unresolved binarity, with the use of binary tracks (shown as dotted 
lines). To construct these binary tracks we started with the single star point which is to 
form the primary. Then, by combining the magnitude of this primary with that of the faintest 
point on the sequence (lowest mass), we obtained a new binary point which we joined to our 
primary point by a dotted line. We then repeated this process for higher mass secondaries, 
joining each new point to its predecessor until we reach an equal mass binary, 0.75 magnitudes 
above the original primary. We also show 5 mass points on the CMDs (in units of 0.001M$_{\odot}$). 
The 0.07M$_{\odot}$ point just above the M dwarf gap is the NextGen model prediction for a 125Myr 
isochrone. Since we believe that the M dwarf gap results from a sharp drop in the luminosity-mass 
relation, we assumed that the mass just below the M dwarf gap is very similar to the mass just 
above it. Hence, we have a 0.07M$_{\odot}$ point just below the M dwarf gap also. The 0.06M$_{\odot}$ 
mass point was estimated from the Pleiades mass-I relation of Dobbie et al. (2002c), which accounts 
for the M dwarf gap, and was derived using the Pleiades I-band BD LF, and an assumed power-law MF. 
We note that this agrees with the NextGen and Dusty model J-band predictions, if they are offset 
by the J magnitude width of the M dwarf gap. The final two mass points are Dusty model predictions 
for a 125Myr isochrone. The 0.04M$_{\odot}$ mass point should be reasonably representative of the 
point where the sequence joins the Dusty isochrone. We also note that Dobbie et al. (2002c) suggest 
that model predictions for the masses of Pleiades BDs from I=18.5-19.5 are too low (by up to 
0.01M$_{\odot}$), which may have some bearing on the Dusty mass estimates.

When analysing these CMDs, we must consider the depth effect of the cluster. Sources near the 
front and back of the cluster will appear slightly brighter and fainter respectively, than they 
would in the cluster centre. The tidal radius of the Pleiades is 13.1pc (Pinfield, Jameson \& 
Hodgkin 1998) giving a full depth of $\pm$0.2$^{mag}$. However, the majority of sources will be 
found closer to the cluster centre. Jameson et al. (2002) fit a King profile (King 1962) to the 
spatial distribution of Pleiades BDs, and derived a core radius of 5pc. We would expect $\sim$80\% 
of members to lie within this core radius, giving a more likely depth of $\pm$0.1$^{mag}$. In order 
to estimate binary mass-ratios (q) for our unresolved binary candidates, we compared their positions 
to the binary tracks in both CMDs, and estimated q ranges consistent with the photometric errors 
and the depth effect of the Pleiades. We did this for a depth effect of $\pm$0.1 and $\pm$0.2. We 
also estimated q ranges on the basis that the Dusty isochrone masses may be higher by 0.01
M$_{\odot}$. And finally, we made a small allowance for the known (or suspected) variables (PPL1 
and Calar 3). Our results are summarised in Table 2. Column 3 indicates if the source is an IK, a 
JK or a JHK binary (identified from the IK CMD, the JK CMD of the JHK diagram). Column 4 gives 
the estimated q range from the mass points in Figure 9 and a $\pm$0.1 depth effect. Column 5 
gives a larger q range, estimated with the full $\pm$0.2 depth effect, and accounting for the 
uncertainty in the Dusty isochrone mass points (assuming they may be 0.01M$_{\odot}$ too low).

Where a source appeared in both CMDs, we found q ranges that were consistent (to within 
photometric errors) for all but 2 candidates; CFHT-PL-18 (BPL 283) and BPL 45 appear red in 
the JK CMD, but do not appear sufficiently red in the IK CMD. The component sources of these 
binaries would therefore be too blue in I-K to be cluster members. This is confirmed for 
CFHT-PL-18, which is a known background binary (see Section 3.4). We have also flagged BPL 45 
as a probable non-member, based on its measured photometry alone. CFHT-PL-12 (BPL 294) and 
INT-PL-IZ-42 could have primaries that are either just above or just below the M dwarf gap. 
However, CFHT-PL-12 would need to be near the front of the cluster if its primary was from below 
the gap. Roque 4 lies quite high in the JK CMD suggesting that it may be a foreground object. 
However, it could still be a binary BD if it is at the front of the cluster, and may alternatively 
be a BD triple system. We were unable to estimate good q ranges for the 2 faintest binaries, 
since we do not know the form of the cluster sequence for masses below 0.03M$_{\odot}$.

It is clear that the majority of the unresolved binary candidates have fairly high q values. 
Indeed, 15 of the 18 probable binary BDs appear to have q$>$0.7. However, the photometry of 3 
sources is inconsistent with them being either single or q$>$0.7 binaries. One of these is 
without J- or H-band measurements, and has not been confirmed as a cluster member (by either 
a proper motion measurement or the lithium test). The other 2 are Teide~1 and Calar~3, which 
are bona fide cluster BDs. We have previously discussed (Section 3.4) other possible causes for 
the red J-K colours seen in these sources, and concluded that binarity was the most likely. We 
note in addition that the position of these sources in the JHK diagram is also consistent with 
binarity, since they appear partway between bins D/E and the location of Roque~25. These few 
objects therefore suggest that BD binaries are not all high q systems. They could consist of 
two populations; the majority in high q systems, and a smaller fraction in low q systems. 
Theoretical simulations are not yet at a level to make such predictions. However, the 
hydrodynamical simulation of Bate, Bonnell \& Bromm (2002) shows BDs forming from the collapse 
and fragmentation of a turbulent molecular cloud. They determined that $\sim$75\% of BDs form 
by the fragmentation of gravitationally unstable circum-stellar disks, with subsequent ejection 
removing the BD from the region of dense gas. The remaining 25\% were found to form from the 
fragmentation of dynamically collapsing filamentary gas, where the forming BDs fell into, and 
were rapidly ejected from unstable multiple systems (see also Reipurth \& Clarke 2001). However, 
this simulation only produced 1 BD binary, but lacked the spatial resolution to predict very close 
systems. Future simulations with higher spatial resolution may reveal a link between the two 
different formation mechanisms and the resulting q distribution of BD binaries.

\subsection{Binary fractions}
\begin{table}
\caption{Binary fractions for the Praesepe and Pleiades samples. Spectral types were 
estimated using Leggett (1992).}
\begin{tabular}{|l|l|l|l|}
\hline
\multicolumn{4}{l}{Praesepe}\\
\hline
I-K & Mass/SpT & q-sensitivity & Binary fraction \\
\hline
1--1.9   & 1.0--0.6M$_{\odot}$               & 0.5--1  & 17$^{+6}_{-4}$\%  \\
         & G/K stars                         &         &             \\
1.9--2.5 & 0.6--$\sim$0.2M$_{\odot}$         & 0.35--1 & 41$\pm$5\%  \\
         & M0--M5                            &         &             \\
1.9--2.2 & 0.6--$\sim$0.35M$_{\odot}$        & 0.3--1  & 31$^{+7}_{-6}$\%   \\
         & M0--M3                            &         &             \\
2.2--2.5 & $\sim$0.35--$\sim$0.2M$_{\odot}$  & 0.4--1  & 44$\pm$6\%  \\
         & M3--M5                            &         &             \\
3--3.6   & $\sim$0.11--$\sim$0.09M$_{\odot}$ & 0.65--1 & 47$^{+13}_{-11}$\% \\
         & M6--M7                            &         &             \\
\hline
\multicolumn{4}{l}{Pleiades}\\
\hline
1--1.9   & 1.0--0.6M$_{\odot}$               & 0.5--1  & 23$^{+6}_{-5}$\%  \\
         & G/K stars                         &         &             \\
2.6--3   & 0.15--0.09M$_{\odot}$             & $\sim$0.35--1 & 36$\pm$5\%  \\
         & M5--M6                            &         &             \\
2.6--2.8 & 0.15--0.11M$_{\odot}$             & 0.35--1 & 27$^{+6}_{-5}$\%  \\
         & M5                                &         &             \\
2.8--3   & 0.11--0.09M$_{\odot}$             & 0.4--1  & 51$\pm$8\% \\
         & M6                                &         &             \\
3.3--4.3 & 0.07--0.06M$_{\odot}$             & 0.5--1  & 50$^{+11}_{-10}$\% \\
         & M7--M8                            &         &             \\
\hline
\end{tabular}
\end{table}

We determined binary fractions (BF; defined as the number of binary systems in some photometric range 
divided by the total number of systems in that range) using the IK CMDs in general, for a number of 
photometric ranges. The results are given in Table 3, where the uncertainties are determined assuming 
binomial statistics (see Burgasser et~al. 2002). The mass ranges were estimated using our IK 
cluster sequences (see Section 3.2). The stellar q ranges were estimated by simulating the effect 
of unresolved binarity in the IK CMDs (as was done in Section 3.7 for the Pleiades BDs), and finding 
the q value when the binary track crosses midway between the single star sequence and the binary 
sequence 0.75 magnitudes above it.

Starting in Praesepe, the first range we considered was from I-K=1--1.9 (G and K stars). The 
single stars all bunch around the predicted sequence, and we flagged binaries by their proximity 
to the binary star sequence. The next range we chose was from I-K=1.9--2.5 (where the upper limit 
represents the completeness of 2MASS and the HSHJ survey). The sources lie in a fairly straight path 
down the CMD in this range, and we therefore chose to separate binaries and single stars using 
a straight line. We defined this separation line such that the average spread of members above 
and below it was the same (as one would expect for a binary induced spread). And by doing this 
for two sub-ranges (I-K=1.9--2.2 and 2.2--2.5) we ensured that the gradient of the line was correct. 
We would expect a spread of $\pm$0.375 to represent the full spread induced from binarity, and we 
obtained spreads of $\pm$0.36 and $\pm$0.43 for the two ranges respectively. The slightly larger 
value for the redder range is accounted for by larger photometric errors. This approach has the added 
benefit that it is independent of any assumed distance modulus. Table 3 gives the binary fraction 
for both sub-ranges, as well as for the full range. For the lowest-mass Praesepe stars we chose to 
use I-K=3--3.6 (where we just have UKIRT photometry). This avoids the CMD overlap region between the 
photographic and CCD surveys, where different survey areas and photometric errors complicate matters. 
Binaries in this range were flagged in Section 3.2. Unlike the higher mass ranges, these candidates 
do not have proper motions. However, using the disk luminosity function of Kirkpatrick et al. (1994), 
we expect no more than $\sim$1 and $\sim$0.5 field star to contaminate the single and binary star 
sequences respectively. Such numbers will not significantly affect our statistics.

For the Pleiades, the first range we considered was from I-K=2.6--3. We took the same approach 
as for Praesepe, and divided this range in two (I-K=2.6--2.8 and 2.8--3), defining a selection line 
such that the scatter above and below (in each region) was identical. The spread about the selection 
line was $\pm$0.57 for each of the sub-regions. This is somewhat more than the expected $\pm$0.375, 
but the additional spread will be predominantly due to photometric errors and cluster depth. Finally, 
we considered a BD range. We counted single BDs below the M dwarf gap that have K$<$15.5, since 
our sample is spatially complete in this range. The mass range for the single stars is $\sim$0.07-0.05
M$_{\odot}$, where we have interpolated between the 0.06 and the 0.04M$_{\odot}$ points in Figure 9. 
We then counted binary BDs that could have primaries amongst these single sources, including objects 
from K=14.2-15.2. We note that we ignored CFHT-PL-12 (BPL 294) since it is likely that its primary 
comes from above the gap. In this way, we selected 10 single BDs, and 10 binary BDs. We estimated the 
q sensitivity range for our BD binaries using Table 2.

We have already discussed how the cluster depth affects BD binaries. However, it will also be 
a source of error for our stellar BFs. Objects whose position on the CMD is within the cluster 
depth (eg. 0.2 and 0.15 magnitudes for the Pleiades and Praesepe respectively) of the binary 
selection line could potentially spill over this line if they were near the front or the back of 
the cluster. This represents 53\% of Pleiades sources, and 40\% of Praesepe sources. However, the 
cluster spatial distributions (Pinfield, Jameson \& Hodgkin 1998; Holland et al. 2000) indicate 
that of these stars, we would only expect $\sim$5-15\% and $\sim$5-10\% of Pleiades and Praesepe 
stars respectively to actually spill across the line. Further more, an error will only occur if 
the BF itself is different from 50\% (for BF=50\% misidentified binaries and single stars will 
cancel out). Combining these together, we would expect no more than $\sim$1\% and 2\% BF errors 
in Praesepe and the Pleiades respectively, due to depth effects.

Another source of error comes from variability, resulting from chromospheric activity such 
as spots or flares pushing some single stars intermittently above the binary selection line.
Lockwood et al. (1984) found that $\sim$40\% of Hyades F G and K dwarfs varied at a level 
of $\sim$1--2\%. For later types, Stauffer et al. (1991) reported H$_{\alpha}$ emission in 
$\sim$60\% of late K--mid M Hyades members, aswell as in the vast majority of G--mid M Pleiades 
members, and Scholz, Eisloffel \& Mundt (2000) found photometric variations among young M-dwarfs 
similar to those of solar type stars. Therefore, although many of our members may be variable,
we expect this variability to be generally low (a few \%). More violent flares are occasionally 
seen in late M dwarfs (Leibert et al. 1999), but are rare. We estimate that our stellar BF values 
should not be increased by more than $\sim$5\% as a result of variability.

\begin{figure}
 \epsfig{file=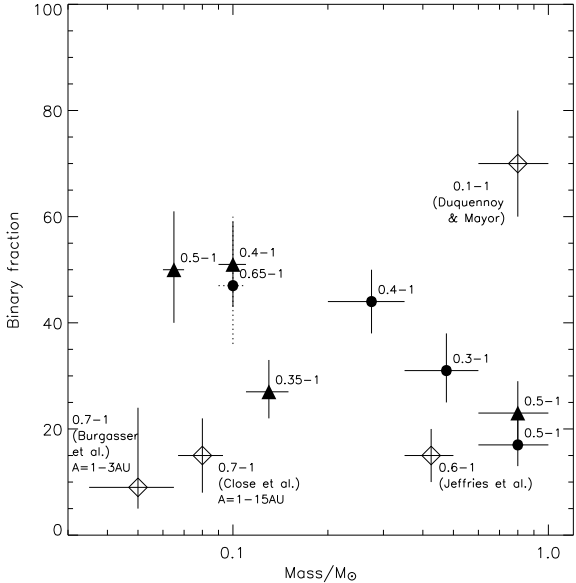,width=8.0cm}
 \caption{The binary fractions of the Pleiades (triangles) and Praesepe (Circles). Three points 
from the literature are shown as open diamonds. At 0.1M$_{\odot}$ the BF error bars of two points 
overlap, so we plot the Praesepe error bars as a dotted line for clarity. q ranges are indicated 
next to each point, and an $A$ range for the Burgasser and Close points.}
\end{figure}

The BF values are plotted in Figure 11, as a function of mass. Filled triangles and circles represent 
Pleiades and Praesepe data respectively. Additional points from the literature are shown as open 
diamonds. The highest mass point agrees well with the field star results of Duquennoy \& Mayor (1991), 
which increases to 70\% for q=0.1--1. From 0.6--0.35M$_{\odot}$ we find a BF of $\sim$30\%. The BF of 
NGC~2516 agrees with this value for $\sim$0.6M$_{\odot}$, but drops to $\sim$15\% for 
0.5--0.35M$_{\odot}$ (Jeffries, Thurston \& Hambly 2001). This difference could be explained by the 
fact that the NGC~2516 CMD was only sensitive to q=0.6--1 binaries, whereas our CMD is sensitive to 
q=0.3--1. If the clusters have intrinsically similar binary populations, then this implies that there 
is a BF of 15\% from q=0.3--0.6 as well as from q=0.6--1, suggesting a flat BF--q distribution in this 
mass range. This represents a departure from the BF--q distribution of solar type field stars, which 
rises towards lower q. If the BF--q distribution were like that of solar type stars we would expect a 
BF of 30\% for q=0.3--0.6, and thus 45\% for q=0.3--1. Our result is 1.5$\sigma$ lower than this value. 
A flat binary--q distribution would give a total BF of $\sim$45\% in this mass range. Next, with the 
exception of the Pleiades point at 0.13M$_{\odot}$, there is a general trend of increasing BF towards 
lower mass. The q ranges of these points are fairly similar, and so should not significantly bias this 
trend. Moving into the sub-stellar regime, it can be seen that the BD binary fraction is around 50\%. 
Even if we are extremely conservative, and only select the most obvious BD binaries from the IK CMD 
(assuming that JK and JHK binaries are in fact single), we find a BF of 32$^{+11}_{-8}$\%. Figure 11 
also shows the T dwarf BF resolved by the HST (Burgasser et~al. 2003), and the late M/early L dwarf 
BF resolved by Hokupa'a on Gemini (Close et~al. 2003). Mass range estimates were taken from Close 
et~al. (2003), and estimated for the T dwarfs assuming an age range of 1-5Gyrs. Both these BFs 
(9$^{+15}_{-4}$\% and 15$\pm$7\% respectively) are significantly lower than our BF for Pleiades BDs. 
However, these observations were only sensitive to binaries with separations $>$1AU. It is possible 
that the total BF of these populations is identical to that of Pleiades BDs if 70-80\% of BD 
binaries have separations $<$1AU. It may be that many BD binaries are like PPL15 (Basri \& Mart\'{\i}n, 
1999) which has a seperation of 0.03AU. We also note that the N-body simulations of Kroupa (2001), 
and Kroupa, Aarseth \& Hurley (2001) suggest that $\sim$50\% of cluster BD binaries with 
$\log{P/days}$=1-6 (seperations of $\sim$0.04--90AU) would be disrupted during the first few Myrs 
of proto-cluster dynamical evolution. This would require an even higher initial BD BF unless most 
BD binaries are extremely tight. However, the hydrodynamical simulation of Bate, Bonnell \& Bromm 
(2002) predicts an initial BF of $\sim$5\%, since only 1 BD binary was produced in their simulation 
among 20 single BDs. But the spatial resolution of this simulation was not high enough to model the 
formation of extremely close BD binaries. Such binaries may form during a secondary collapse phase 
of individual pressure supported fragments (Bonnell \& Bate 1994). If this is not the case (Bate 1998), 
then post fragmentation processes (such as dynamical interactions) might be responsible for creating 
extremely tight BD binaries (by hardening).

\section{Future work}
Proper motion measurements have already been made for 15 of the BD candidates discussed here. 
Work is currently underway by some of us to measure proper motions of the remaining Pleiades 
BDs. It will be particularly useful to confirm all possible low-q BD binaries as Pleiades members. 
Measuring proper motions of the Praesepe candidates is also important, and is in progress (Chappelle, 
Pinfield \& Steele 2002). Such measurements will allow us to define the Praesepe cluster sequence 
with more confidence. Also, the faintest Praesepe candidates could be BDs, and would be the first 
identified in an intermediate age cluster.

Measuring NIR spectra is a natural next step if we are to understand the nature of the cluster 
sequences. These spectra could be compared to the predictions of both NextGen and Dusty model 
atmospheres, to determine when dust is beginning to form in the atmospheres of Pleiades and Praesepe 
sources, and what effects it has in the different surface gravities. Furthermore, the best way to 
improve on our analysis of binarity is from NIR spectra. We could combine the spectra of likely 
single sources, and compare the results with spectra of suspected unresolved binaries, to confirm 
or not, if these sources really are binary in nature.

The nature of the M-dwarf gap could be further investigated by measuring longer wavelength photometry 
(eg. L-band) of sources on either side of the M dwarf gap, to look for evidence (or lack of it) of 
dust formation.

Deeper and larger scale surveys of the clusters will provide much larger samples of members 
to work with. The 16 square degree survey recently obtained as part of the WFS with the INT and 
WFC should identify several hundred Pleiades BDs. Also, the WFC survey of Praesepe made by Chappelle, 
Pinfield \& Steele (2002) should identify large numbers of very low-mass stars and possibly BDs. With 
these larger samples we will be able to more accurately measure the M dwarf gap in both clusters, 
as well as obtain a better statistical measure of the BD binary fraction and binary-q distribution.

\section*{Acknowledgments}

DJP and PDD acknowledge funding from PPARC. AK acknowledges funding from the Leverhulme Trust.
The United Kingdom Infrared Telescope is operated by the Joint Astronomy Centre on behalf of 
the U.K. Particle Physics and Astronomy Research Council. Some of the data reported here were 
obtained as part of the UKIRT Service Program. This publication makes use of data products 
from the Two Micron All Sky Survey, which is a joint project of the University of Massachusetts 
and the Infrared Processing and Analysis Center/California Institute of Technology, funded by 
the Aeronautics and Space Administration and the National Science Foundation.

\appendix

\section[]{Tables of photometry}

\begin{table*}
\begin{center}
\caption{Photometry (c3-6), membership criterea (c7-9,c11-13) and binarity (c10) of Pleiades 
BD candidates. Membership is from the IK CMD, the JK CMD and the JHK diagram. 
Refs; (a) Mart\'{\i}n, Rebolo \& Zapatero-Osorio (1996), (b) Rebolo et al. (1996), (c) Zapatero Osorio et al. (1997), 
(d) Mart\'{\i}n et al. (1998) (e) Stauffer et al. (1998), (f) Mart\'{\i}n et al. (2000), (g) Pinfield 
et al. (2000), (h) Moraux, Bouvier \& Stauffer (2001), (i) Dobbie et al. (2002), (j) Dobbie 
(2002) private comm. A $^{\star}$ indicates a non-member by virtue of binarity.}
\begin{tabular}{|l|l|l|l|l|l|l|l|l|l|l|l|l|l|}
\hline
Name & Other names & I$_C$ & J$_{MKO}$ & H$_{MKO}$ & K$_{MKO}$ & {\it IK} & {\it JK} & {\it JHK} & Bin & SpT & pm & Li \\
\hline
BPL 45  &             & 18.61$^g$ & 16.07$\pm$0.05     & 15.42$\pm$0.03     & 14.93$\pm$0.04     & y & N$^{\star}$ & y & b & y &    &           \\
BPL 49  & CFHT-Pl-17  & 18.55$^g$ & 16.12$\pm$0.04$^g$ & 15.49$\pm$0.07$^g$ & 15.07$\pm$0.06$^g$ & y & y & y &   &       & y$^h$  &           \\
        &             &           & 16.14$\pm$0.10$^i$ & 15.51$\pm$0.11$^i$ &                    &   &   &   &   &       &        &           \\
        &             &           & 15.99$\pm$0.04     & 15.48$\pm$0.05     &                    &   &   &   &   &       &        &           \\
BPL 62  & Roque 7     & 19.44$^g$ & 16.53$\pm$0.03$^g$ & 15.85$\pm$0.02$^g$ & 15.47$\pm$0.02$^g$ & y & y & y &   &       & y$^h$  &           \\
        & CFHT-Pl-24  &           & 16.55$\pm$0.18$^i$ & 15.88$\pm$0.14$^i$ &                    &   &   &   &   &       &        &           \\
BPL 66  & Roque 4     & 19.81$^g$ & 16.66$\pm$0.05     & 15.93$\pm$0.03     & 15.26$\pm$0.04     & y & y & y & b & y$^c$ &        &           \\
        &             &           & 16.62$\pm$0.04     & 15.92$\pm$0.05     & 15.26$\pm$0.04     &   &   &   &   &       &        &           \\
        &             &           & 16.69$\pm$0.12$^i$ & 16.04$\pm$0.09$^i$ &                    &   &   &   &   &       &        &           \\
BPL 76  &             & 17.83$^g$ & 15.41$\pm$0.04     & 14.85$\pm$0.03     & 14.57$\pm$0.06$^g$ & y & y & y &   &       &        &           \\
BPL 78  &             & 18.98$^g$ & 16.02$\pm$0.05     & 15.45$\pm$0.03     & 15.01$\pm$0.04     & y & y & y &   &       &        &           \\
BPL 79  & Roque 13    & 18.69$^g$ & 15.63$\pm$0.03$^g$ & 15.09$\pm$0.02$^g$ & 14.56$\pm$0.03$^g$ & y & y & y & b & y$^c$ &        &           \\
        &             &           & 15.65$\pm$0.09$^i$ & 15.11$\pm$0.08$^i$ &                    &   &   &   &   &       &        &           \\
BPL 81  &             & 18.79$^g$ & 15.97$\pm$0.03     & 15.39$\pm$0.05     & 14.98$\pm$0.02$^g$ & y & y & y &   &       &        &           \\
BPL 100 & Roque 9     & 19.33$^g$ & 16.28$\pm$0.03     & 15.63$\pm$0.05     & 15.22$\pm$0.02$^g$ & y & y & y &   &       &        &           \\
        &             &           & 16.30$\pm$0.19$^i$ & 15.72$\pm$0.15$^i$ &                    &   &   &   &   &       &        &           \\
BPL 108 & Roque 14    & 18.53$^g$ & 15.50$\pm$0.02$^g$ & 14.89$\pm$0.02$^g$ & 14.47$\pm$0.03$^g$ & y & y & y & b & y$^c$ &        &           \\
        &             &           & 15.53$\pm$0.10$^i$ & 14.96$\pm$0.09$^i$ &                    &   &   &   &   &       &        &           \\
        &             &           & 15.59$\pm$0.03     & 14.96$\pm$0.03     &                    &   &   &   &   &       &        &           \\
BPL 124 &             & 17.78$^g$ & 15.43$\pm$0.05     & 14.85$\pm$0.03     & 14.46$\pm$0.04     & y & y & y &   &       & y$^g$  &           \\
BPL 132 & Roque 11    & 19.06$^g$ & 16.08$\pm$0.03$^g$ & 15.64$\pm$0.03$^g$ & 15.13$\pm$0.04$^g$ & y & y & N &   & y$^b$ &        &           \\
        & NPL 37      &           &                    &                    &                    &   &   &   &   &       &        &           \\
BPL 137 & Teide 1     & 19.22$^g$ & 16.32$\pm$0.05$^g$ & 15.62$\pm$0.05$^g$ & 15.08$\pm$0.05$^g$ & y & y & y & b & y$^b$ & y$^h$  & y$^c$     \\
        & NPL 39      &           & 16.19$\pm$0.11$^i$ & 15.65$\pm$0.09$^i$ &                    &   &   &   &   &       &        &           \\
        &             &           & 16.24$\pm$0.05     & 15.62$\pm$0.05     &                    &   &   &   &   &       &        &           \\
BPL 142 & Roque 17    & 18.11$^g$ & 15.29$\pm$0.03$^g$ & 14.77$\pm$0.03$^g$ & 14.36$\pm$0.05$^g$ & y & y & y & b & y$^c$ & y$^g$  &           \\
        &             &           & 15.31$\pm$0.09$^i$ & 14.79$\pm$0.07$^i$ &                    &   &   &   &   &       &        &           \\
BPL 152 & Roque 16    & 17.95$^g$ & 15.49$\pm$0.03$^g$ & 14.97$\pm$0.02$^g$ & 14.62$\pm$0.03$^g$ & y & y & y &   & y$^c$ & y$^h$  & y$^e$     \\
        & CFHT-Pl-11  &           & 15.51$\pm$0.11$^i$ & 14.99$\pm$0.09$^i$ &                    &   &   &   &   &       &        &           \\
BPL 163 &             & 17.83$^g$ & 15.31$\pm$0.03     & 14.78$\pm$0.05     & 14.53$\pm$0.04$^g$ & y & N & y &   &       &        &           \\
BPL 168 &             & 17.65$^g$ & --                 & --                 & 16.02$\pm$0.10     & N & --& --&   &       &        &           \\
BPL 172 & Roque 12    & 18.53$^g$ & 15.91$\pm$0.03     & 15.36$\pm$0.05     & 15.10$\pm$0.03$^g$ & y & y & y &   & y$^c$ &        &           \\
        & NPL 36      &           & 15.93$\pm$0.10$^i$ & 15.44$\pm$0.08$^i$ & 14.97$\pm$0.07$^i$ &   &   &   &   &       &        &           \\
BPL 201 &             & 19.60$^g$ & 16.40$\pm$0.03$^g$ & 15.75$\pm$0.03$^g$ & 15.35$\pm$0.04$^g$ & y & y & y &   &       &        &           \\
BPL 202 & CFHT-Pl-9   & 17.75$^g$ & 15.44$\pm$0.03     & 14.88$\pm$0.05     & 14.49$\pm$0.04$^g$ & y & y & y &   & y$^e$ & y$^h$  & N$^e$     \\
        &             &           & 15.40$\pm$0.04     & 14.84$\pm$0.02     &                    &   &   &   &   &       &        &           \\
BPL 235 & Calar 3     & 18.91$^g$ & 16.24$\pm$0.03$^g$ & 15.53$\pm$0.03$^g$ & 14.91$\pm$0.03$^g$ & y & y & y & b & y$^a$ & y$^h$  & y$^b$     \\
        & CFHT-Pl-21  &           & 16.09$\pm$0.04     & 15.49$\pm$0.05     &                    &   &   &   &   &       &        &           \\
BPL 240 &             & 18.68$^g$ & 15.99$\pm$0.03     & 15.44$\pm$0.05     & 15.11$\pm$0.03$^g$ & y & y & y &   &       &        &           \\
BPL 249 &             & 20.02$^g$ & 16.66$\pm$0.03$^g$ & 16.13$\pm$0.03$^g$ & 15.46$\pm$0.03$^g$ & y & y & y & b &       &        &           \\
BPL 254 & Teide 2     & 17.81$^g$ & 15.53$\pm$0.04     & 14.94$\pm$0.03     & 14.57$\pm$0.06$^g$ & y & y & y &   & y$^d$ & y$^h$  & y$^{d,e}$ \\
        & CFHT-Pl-13  &           & 15.53$\pm$0.08$^i$ & 14.97$\pm$0.07$^i$ & 14.54$\pm$0.06$^i$ &   &   &   &   &       &        &           \\
BPL 283 & CFHT-Pl-18  & 18.58$^g$ & 16.04$\pm$0.03$^g$ & 15.43$\pm$0.02$^g$ & 14.94$\pm$0.06$^g$ & y & N$^{\star}$ & y & b & y$^f$ & N$^h$ & N$^f$ \\
BPL 294 & CFHT-Pl-12  & 17.85$^g$ & 15.16$\pm$0.03     & 14.55$\pm$0.02     & 14.20$\pm$0.02$^g$ & y & y & y & b & y$^e$ & y$^h$  & y$^e$     \\
BPL 303 & CFHT-Pl-25  & 19.69$^g$ & 16.63$\pm$0.04     & 16.01$\pm$0.02     & 15.47$\pm$0.06$^g$ & y & y & y & b &       & y$^h$  &           \\
        &             &           & 16.64$\pm$0.15$^i$ & 16.05$\pm$0.13$^i$ &                    &   &   &   &   &       &        &           \\
BPL 306 &             & 19.09$^g$ & 16.16$\pm$0.04     & 15.53$\pm$0.02     & 15.15$\pm$0.02$^g$ & y & y & y &   &       &        &           \\
BPL 316 &             & 18.30$^g$ & 15.97$\pm$0.04     & 15.41$\pm$0.02     & 15.01$\pm$0.03     & y & y & y &   &       &        &           \\
        &             &           &                    &                    & 14.93$\pm$0.04     &   &   &   &   &       &        &           \\
BPL 327 & IPMBD 11    & 17.94$^g$ & 15.52$\pm$0.03     & 14.99$\pm$0.02     & 14.60$\pm$0.02$^g$ & y & y & y &   & y     & y$^g$  &           \\
\hline												  	          		   
PPL 1      & Roque 15 & 18.21$^i$ & 15.37$\pm$0.08$^i$ & 14.82$\pm$0.06$^i$ & 14.34$\pm$0.05$^i$ & y & y & y & b & y     &        & y         \\
PPL 15     & IPMBD 23 & 18.24$^i$ & 15.34$^i$          & 14.77$^i$          & 14.41$^i$          & y & y & y & b & y     & y      & y         \\
           & NPL 35   &           &                    &                    &                    &   &   &   &   &       &        &           \\
CFHT-PL-23 &          & 19.33$^i$ & 16.38$\pm$0.08$^i$ & 15.79$\pm$0.08$^i$ & 15.25$\pm$0.05$^i$ & y & y & y & b &       &        &           \\
Roque 5    &          & 20.19$^i$ & 16.89$\pm$0.14$^i$ & 16.33$\pm$0.15$^i$ & 15.74$\pm$0.10$^i$ & y & y & y &   &       &        &           \\
Roque 8    &          & 19.55$^i$ & 16.68$\pm$0.11$^i$ & 16.14$\pm$0.10$^i$ & 15.57$\pm$0.06$^i$ & y & y & y &   &       &        &           \\
Roque 25   &          & 21.80$^i$ & 17.7$\pm$0.3$^i$   & 17.0$\pm$0.3$^i$   & 16.31$\pm$0.19$^i$ & y & y & y &   & y     &        &           \\
Roque 30   &          & 20.82$^i$ & 17.49$\pm$0.12$^i$ & 16.77$\pm$0.11$^i$ & 16.14$\pm$0.07$^i$ & y & y & y &   &       &        &           \\
Roque 33   & NPL 40   & 20.44$^i$ & 17.07$\pm$0.12$^i$ & 16.62$\pm$0.18$^i$ & 16.02$\pm$0.08$^i$ & y & y & ? &   & y     & y$^j$  & y         \\
Roque 36   &          & 20.25$^i$ & 17.17$\pm$0.20$^i$ & 16.58$\pm$0.12$^i$ & 16.09$\pm$0.08$^i$ & y & y & y &   &       &        &           \\
\hline
\end{tabular}
\end{center}
\end{table*}

\begin{table*}
\begin{center}
\caption{Photometry (c2-5) and membership (c6-8) of Pleiades low-mass star candidates 
(higher mass BPL candidates). Membership is from the IK CMD, the JK CMD and the JHK 
diagram. J- H- and K-band photometry is from 2MASS, and has been transformed into the 
MKO system.}
\begin{tabular}{|l|l|l|l|l|l|l|l|}
\hline
BPL & I$_C$ & J$_{MKO}$ & H$_{MKO}$ & K$_{MKO}$ & {\it IK} & {\it JK} & {\it JHK} \\
\hline
  3 & 14.81 & 13.25$\pm$0.03 & 12.74$\pm$0.03 & 12.36$\pm$0.03 & y & y & y \\
  4 & 14.17 & 12.67$\pm$0.03 & 12.12$\pm$0.02 & 11.77$\pm$0.03 & y & y & y \\
  5 & 14.25 & 12.66$\pm$0.03 & 12.17$\pm$0.03 & 11.80$\pm$0.03 & y & y & y \\
  6 & 15.74 & 14.01$\pm$0.03 & 13.53$\pm$0.03 & 13.09$\pm$0.04 & y & y & y \\
  7 & 16.28 & 14.44$\pm$0.03 & 13.92$\pm$0.03 & 13.48$\pm$0.04 & y & y & y \\
 16 & 17.02 & 15.20$\pm$0.04 & 14.67$\pm$0.05 & 14.24$\pm$0.06 & N & y & y \\
 18 & 14.39 & 12.89$\pm$0.03 & 12.35$\pm$0.03 & 11.97$\pm$0.03 & y & y & y \\
 19 & 16.39 & 14.60$\pm$0.03 & 14.09$\pm$0.04 & 13.76$\pm$0.04 & y & y & y \\
 20 & 14.81 & 13.21$\pm$0.03 & 12.70$\pm$0.03 & 12.28$\pm$0.03 & y & y & y \\
 26 & 17.00 & 14.67$\pm$0.03 & 14.16$\pm$0.04 & 13.76$\pm$0.05 & y & y & y \\
 30 & 15.90 & 14.15$\pm$0.03 & 13.58$\pm$0.03 & 13.22$\pm$0.04 & y & y & y \\
 34 & 13.72 & 12.01$\pm$0.02 & 11.45$\pm$0.03 & 11.10$\pm$0.03 & y & y & y \\
 38 & 16.12 & 14.27$\pm$0.03 & 13.78$\pm$0.04 & 13.42$\pm$0.04 & y & y & y \\
 43 & 15.81 & 14.11$\pm$0.03 & 13.56$\pm$0.03 & 13.18$\pm$0.04 & y & y & y \\
 44 & 16.09 & 14.30$\pm$0.03 & 13.76$\pm$0.03 & 13.37$\pm$0.04 & y & y & y \\
 56 & 14.90 & 13.29$\pm$0.03 & 12.74$\pm$0.02 & 12.40$\pm$0.03 & y & y & y \\
 58 & 17.69 & 15.37$\pm$0.05 & 14.77$\pm$0.06 & 14.23$\pm$0.07 & y & y & y \\
 61 & 14.86 & 13.11$\pm$0.02 & 12.58$\pm$0.03 & 12.23$\pm$0.03 & y & y & y \\
 68 & 14.33 & 12.85$\pm$0.03 & 12.34$\pm$0.03 & 12.03$\pm$0.04 & y & y & y \\
 69 & 17.59 & 16.61$\pm$0.15 & 16.10$\pm$0.20 & 15.56$\pm$0.18 & N & y & y \\
 70 & 15.40 & 13.33$\pm$0.02 & 12.76$\pm$0.03 & 12.45$\pm$0.03 & y & y & y \\
 71 & 15.14 & 13.18$\pm$0.02 & 12.58$\pm$0.03 & 12.23$\pm$0.03 & y & y & y \\
 72 & 16.18 & 14.04$\pm$0.03 & 13.52$\pm$0.03 & 13.20$\pm$0.03 & y & y & y \\
 73 & 15.09 & 13.08$\pm$0.02 & 12.60$\pm$0.03 & 12.21$\pm$0.03 & y & y & y \\
 74 & 17.44 & 15.19$\pm$0.05 & 14.68$\pm$0.06 & 14.23$\pm$0.06 & y & y & y \\
 75 & 16.28 & 14.03$\pm$0.03 & 13.46$\pm$0.03 & 13.06$\pm$0.03 & y & y & y \\
 84 & 17.21 & 14.81$\pm$0.04 & 14.29$\pm$0.04 & 13.85$\pm$0.04 & y & y & y \\
 89 & 13.89 & 12.01$\pm$0.03 & 11.46$\pm$0.03 & 11.14$\pm$0.03 & y & y & y \\
 91 & 14.74 & 12.73$\pm$0.03 & 12.21$\pm$0.03 & 11.88$\pm$0.03 & y & y & y \\
 92 & 15.44 & 13.32$\pm$0.03 & 12.83$\pm$0.03 & 12.44$\pm$0.03 & y & y & y \\
 93 & 14.63 & 12.71$\pm$0.03 & 12.18$\pm$0.03 & 11.81$\pm$0.03 & y & y & y \\
 94 & 15.26 & 13.58$\pm$0.03 & 13.08$\pm$0.03 & 12.74$\pm$0.03 & y & y & y \\
 95 & 14.02 & 12.51$\pm$0.03 & 11.96$\pm$0.03 & 11.65$\pm$0.03 & y & y & y \\
 96 & 16.38 & 14.05$\pm$0.03 & 13.51$\pm$0.04 & 13.13$\pm$0.03 & y & y & y \\
 97 & 14.55 & 12.82$\pm$0.03 & 12.32$\pm$0.03 & 11.96$\pm$0.03 & y & y & y \\
 98 & 15.82 & 13.95$\pm$0.03 & 13.39$\pm$0.03 & 13.11$\pm$0.04 & y & y & y \\
 99 & 15.30 & 13.60$\pm$0.03 & 13.08$\pm$0.03 & 12.74$\pm$0.03 & y & y & y \\
101 & 16.70 & 14.74$\pm$0.04 & 14.19$\pm$0.05 & 13.82$\pm$0.06 & y & y & y \\
102 & 15.11 & 13.40$\pm$0.03 & 12.87$\pm$0.03 & 12.55$\pm$0.03 & y & y & y \\
103 & 14.57 & 12.90$\pm$0.03 & 12.38$\pm$0.03 & 12.05$\pm$0.03 & y & y & y \\
104 & 16.26 & 14.09$\pm$0.03 & 13.54$\pm$0.04 & 13.19$\pm$0.03 & y & y & y \\
105 & 15.49 & 13.33$\pm$0.03 & 12.84$\pm$0.03 & 12.43$\pm$0.03 & y & y & y \\
106 & 17.44 & 15.22$\pm$0.06 & 14.62$\pm$0.07 & 14.15$\pm$0.07 & y & y & y \\
107 & 16.18 & 14.04$\pm$0.03 & 13.54$\pm$0.04 & 13.16$\pm$0.04 & y & y & y \\
109 & 16.02 & 14.18$\pm$0.03 & 13.70$\pm$0.04 & 13.28$\pm$0.05 & y & y & y \\
110 & 16.70 & 14.40$\pm$0.03 & 13.83$\pm$0.04 & 13.47$\pm$0.05 & y & y & y \\
111 & 15.98 & 14.19$\pm$0.03 & 13.62$\pm$0.04 & 13.23$\pm$0.04 & y & y & y \\
112 & 17.30 & 14.98$\pm$0.05 & 14.37$\pm$0.06 & 14.21$\pm$0.08 & y & y & y \\
113 & 13.73 & 12.02$\pm$0.02 & 11.45$\pm$0.03 & 11.12$\pm$0.02 & y & y & y \\
114 & 13.85 & 12.34$\pm$0.02 & 11.82$\pm$0.03 & 11.48$\pm$0.03 & y & y & y \\
115 & 17.00 & 14.90$\pm$0.05 & 14.37$\pm$0.05 & 13.98$\pm$0.06 & y & y & y \\
116 & 15.79 & 13.75$\pm$0.03 & 13.27$\pm$0.03 & 12.87$\pm$0.03 & y & y & y \\
117 & 14.64 & 13.00$\pm$0.03 & 12.49$\pm$0.03 & 12.12$\pm$0.03 & y & y & y \\
118 & 14.82 & 13.07$\pm$0.02 & 12.53$\pm$0.03 & 12.17$\pm$0.03 & y & y & y \\
119 & 15.10 & 13.23$\pm$0.03 & 12.67$\pm$0.03 & 12.29$\pm$0.04 & y & y & y \\
120 & 15.06 & 13.11$\pm$0.03 & 12.59$\pm$0.03 & 12.23$\pm$0.03 & y & y & y \\
121 & 15.03 & 13.20$\pm$0.03 & 12.67$\pm$0.03 & 12.33$\pm$0.03 & y & y & y \\
\hline
\end{tabular}
\end{center}
\end{table*}

\begin{table*}
\begin{center}
\begin{tabular}{|l|l|l|l|l|l|l|l|}
\hline
BPL & I$_C$ & J$_{MKO}$ & H$_{MKO}$ & K$_{MKO}$ & {\it IK} & {\it JK} & {\it JHK} \\
\hline
122 & 17.13 & 14.96$\pm$0.05 & 14.46$\pm$0.06 & 14.04$\pm$0.07 & y & y & y \\
123 & 17.32 & 15.74$\pm$0.08 & 14.98$\pm$0.09 & 14.90$\pm$0.15 & N & y & N \\
125 & 15.80 & 13.98$\pm$0.03 & 13.40$\pm$0.04 & 13.08$\pm$0.03 & y & y & y \\
126 & 14.34 & 12.74$\pm$0.02 & 12.18$\pm$0.03 & 11.87$\pm$0.03 & y & y & y \\
127 & 17.71 & 15.37$\pm$0.06 & 14.84$\pm$0.07 & 14.32$\pm$0.08 & y & y & y \\
128 & 16.45 & 14.44$\pm$0.03 & 13.91$\pm$0.04 & 13.64$\pm$0.05 & y & y & y \\
129 & 16.23 & 14.23$\pm$0.03 & 13.76$\pm$0.04 & 13.43$\pm$0.04 & y & y & y \\
130 & 17.27 & 14.81$\pm$0.04 & 14.27$\pm$0.05 & 13.97$\pm$0.06 & y & y & y \\
131 & 16.12 & 14.00$\pm$0.03 & 13.55$\pm$0.04 & 13.18$\pm$0.04 & y & y & y \\
133 & 14.02 & 12.34$\pm$0.03 & 11.80$\pm$0.03 & 11.45$\pm$0.03 & y & y & y \\
134 & 15.60 & 13.62$\pm$0.02 & 13.12$\pm$0.03 & 12.75$\pm$0.03 & y & y & y \\
135 & 15.22 & 13.21$\pm$0.04 & 12.73$\pm$0.03 & 12.36$\pm$0.03 & y & y & y \\
136 & 15.29 & 13.37$\pm$0.03 & 12.84$\pm$0.03 & 12.46$\pm$0.04 & y & y & y \\
138 & 15.09 & 13.14$\pm$0.02 & 12.65$\pm$0.03 & 12.29$\pm$0.03 & y & y & y \\
139 & 16.23 & 14.07$\pm$0.03 & 13.53$\pm$0.03 & 13.17$\pm$0.04 & y & y & y \\
140 & 16.44 & 14.22$\pm$0.04 & 13.71$\pm$0.04 & 13.32$\pm$0.04 & y & y & y \\
141 & 14.27 & 12.64$\pm$0.03 & 12.13$\pm$0.03 & 11.78$\pm$0.03 & y & y & y \\
143 & 15.70 & 13.84$\pm$0.03 & 13.34$\pm$0.03 & 13.00$\pm$0.03 & y & y & y \\
144 & 14.06 & 12.30$\pm$0.02 & 11.75$\pm$0.03 & 11.41$\pm$0.03 & y & y & y \\
145 & 15.34 & 13.68$\pm$0.03 & 13.17$\pm$0.03 & 12.86$\pm$0.04 & y & y & y \\
146 & 16.77 & 14.97$\pm$0.05 & 14.40$\pm$0.06 & 14.21$\pm$0.07 & N & y & y \\
147 & 15.50 & 13.69$\pm$0.03 & 13.22$\pm$0.03 & 12.87$\pm$0.03 & y & y & y \\
148 & 14.93 & 12.97$\pm$0.04 & 12.42$\pm$0.03 & 12.07$\pm$0.03 & y & y & y \\
149 & 15.45 & 13.67$\pm$0.03 & 13.10$\pm$0.03 & 12.75$\pm$0.04 & y & y & y \\
150 & 15.46 & 13.64$\pm$0.03 & 13.12$\pm$0.03 & 12.74$\pm$0.03 & y & y & y \\
151 & 14.04 & 12.55$\pm$0.03 & 12.03$\pm$0.03 & 11.68$\pm$0.03 & y & y & y \\
153 & 14.76 & 13.03$\pm$0.03 & 12.49$\pm$0.03 & 12.13$\pm$0.03 & y & y & y \\
154 & 16.94 & 14.72$\pm$0.04 & 14.20$\pm$0.05 & 13.78$\pm$0.05 & y & y & y \\
155 & 15.52 & 13.49$\pm$0.03 & 12.88$\pm$0.03 & 12.54$\pm$0.03 & y & y & y \\
156 & 15.30 & 13.41$\pm$0.03 & 12.91$\pm$0.03 & 12.56$\pm$0.04 & y & y & y \\
157 & 14.78 & 12.95$\pm$0.03 & 12.42$\pm$0.03 & 12.07$\pm$0.03 & y & y & y \\
158 & 15.27 & 13.48$\pm$0.03 & 13.01$\pm$0.03 & 12.62$\pm$0.03 & y & y & y \\
159 & 13.78 & 12.30$\pm$0.03 & 11.77$\pm$0.03 & 11.45$\pm$0.03 & y & y & y \\
160 & 15.06 & 12.96$\pm$0.03 & 12.47$\pm$0.04 & 12.12$\pm$0.03 & y & y & y \\
161 & 15.78 & 13.72$\pm$0.04 & 13.20$\pm$0.04 & 12.87$\pm$0.03 & y & y & y \\
162 & 16.07 & 14.15$\pm$0.04 & 13.62$\pm$0.05 & 13.25$\pm$0.05 & y & y & y \\
164 & 16.38 & 14.43$\pm$0.04 & 13.96$\pm$0.05 & 13.50$\pm$0.06 & y & y & y \\
165 & 14.04 & 11.81$\pm$0.03 & 11.37$\pm$0.04 & 10.97$\pm$0.03 & y & y & y \\
166 & 14.09 & 12.35$\pm$0.04 & 11.82$\pm$0.03 & 11.48$\pm$0.03 & y & y & y \\
167 & 13.74 & 12.13$\pm$0.03 & 11.55$\pm$0.05 & 11.24$\pm$0.02 & y & y & y \\
169 & 17.60 & 15.32$\pm$0.06 & 14.71$\pm$0.07 & 14.43$\pm$0.08 & y & y & y \\
170 & 15.68 & 13.89$\pm$0.04 & 13.37$\pm$0.05 & 12.97$\pm$0.03 & y & y & y \\
171 & 15.43 & 13.41$\pm$0.04 & 12.90$\pm$0.03 & 12.59$\pm$0.03 & y & y & y \\
173 & 14.16 & 12.42$\pm$0.03 & 11.92$\pm$0.06 & 11.58$\pm$0.02 & y & y & y \\
174 & 15.55 & 13.72$\pm$0.04 & 13.26$\pm$0.04 & 12.92$\pm$0.03 & y & y & y \\
175 & 14.05 & 12.31$\pm$0.03 & 11.77$\pm$0.03 & 11.48$\pm$0.03 & y & y & y \\
176 & 14.42 & 12.87$\pm$0.04 & 12.39$\pm$0.04 & 12.00$\pm$0.04 & y & y & y \\
177 & 17.06 & 14.86$\pm$0.05 & 14.40$\pm$0.05 & 14.00$\pm$0.06 & y & y & y \\
178 & 14.65 & 12.82$\pm$0.03 & 12.31$\pm$0.03 & 11.93$\pm$0.03 & y & y & y \\
179 & 15.61 & 15.07$\pm$0.05 & 14.77$\pm$0.07 & 14.56$\pm$0.09 & N & N & N \\
180 & 13.72 & 12.20$\pm$0.03 & 11.65$\pm$0.03 & 11.34$\pm$0.03 & y & y & y \\
181 & 17.33 & 14.73$\pm$0.04 & 14.16$\pm$0.05 & 13.70$\pm$0.05 & y & y & y \\
182 & 16.06 & 14.05$\pm$0.03 & 13.52$\pm$0.04 & 13.18$\pm$0.04 & y & y & y \\
183 & 16.84 & 14.85$\pm$0.04 & 14.33$\pm$0.05 & 13.84$\pm$0.05 & y & y & y \\
184 & 14.21 & 12.68$\pm$0.03 & 12.14$\pm$0.03 & 11.81$\pm$0.03 & y & y & y \\
185 & 14.86 & 13.18$\pm$0.04 & 12.66$\pm$0.06 & 12.32$\pm$0.03 & y & y & y \\
186 & 16.59 & 14.57$\pm$0.04 & 14.06$\pm$0.05 & 13.56$\pm$0.04 & y & y & y \\
\hline
\end{tabular}
\end{center}
\end{table*}

\begin{table*}
\begin{center}
\begin{tabular}{|l|l|l|l|l|l|l|l|}
\hline
BPL & I$_C$ & J$_{MKO}$ & H$_{MKO}$ & K$_{MKO}$ & {\it IK} & {\it JK} & {\it JHK} \\
\hline
187 & 14.09 & 12.55$\pm$0.03 & 12.02$\pm$0.03 & 11.67$\pm$0.02 & y & y & y \\
188 & 16.26 & 14.31$\pm$0.04 & 13.80$\pm$0.06 & 13.44$\pm$0.04 & y & y & y \\
189 & 15.22 & 13.45$\pm$0.03 & 12.98$\pm$0.03 & 12.57$\pm$0.03 & y & y & y \\
190 & 15.86 & 14.02$\pm$0.04 & 13.50$\pm$0.06 & 13.11$\pm$0.04 & y & y & y \\
191 & 17.74 & 15.06$\pm$0.05 & 14.51$\pm$0.06 & 14.08$\pm$0.06 & y & y & y \\
192 & 17.46 & 15.09$\pm$0.05 & 14.56$\pm$0.06 & 14.08$\pm$0.06 & y & y & y \\
193 & 13.73 & 12.10$\pm$0.03 & 11.55$\pm$0.03 & 11.27$\pm$0.02 & y & y & y \\
194 & 15.89 & 13.93$\pm$0.04 & 13.45$\pm$0.04 & 13.09$\pm$0.04 & y & y & y \\
197 & 16.42 & 14.17$\pm$0.04 & 13.66$\pm$0.04 & 13.33$\pm$0.04 & y & y & y \\
198 & 16.95 & 14.71$\pm$0.04 & 14.16$\pm$0.05 & 13.74$\pm$0.05 & y & y & y \\
206 & 16.43 & 14.16$\pm$0.04 & 13.63$\pm$0.04 & 13.30$\pm$0.05 & y & y & y \\
208 & 15.30 & 13.34$\pm$0.04 & 12.79$\pm$0.04 & 12.48$\pm$0.03 & y & y & y \\
209 & 14.09 & 12.58$\pm$0.03 & 12.05$\pm$0.03 & 11.73$\pm$0.03 & y & y & y \\
210 & 16.63 & 14.57$\pm$0.04 & 14.08$\pm$0.05 & 13.66$\pm$0.05 & y & y & y \\
211 & 15.99 & 13.88$\pm$0.03 & 13.38$\pm$0.04 & 13.00$\pm$0.04 & y & y & y \\
212 & 14.90 & 13.22$\pm$0.03 & 12.69$\pm$0.03 & 12.37$\pm$0.03 & y & y & y \\
214 & 17.14 & 14.70$\pm$0.04 & 14.10$\pm$0.04 & 13.69$\pm$0.05 & y & y & y \\
215 & 17.10 & 14.99$\pm$0.05 & 14.47$\pm$0.06 & 14.14$\pm$0.06 & y & y & y \\
216 & 14.76 & 12.91$\pm$0.03 & 12.26$\pm$0.04 & 11.97$\pm$0.03 & y & y & y \\
217 & 13.85 & 12.33$\pm$0.03 & 11.83$\pm$0.04 & 11.52$\pm$0.03 & y & y & y \\
218 & 17.38 & 15.13$\pm$0.05 & 14.57$\pm$0.06 & 14.36$\pm$0.09 & y & y & y \\
219 & 16.14 & 14.18$\pm$0.04 & 13.74$\pm$0.04 & 13.37$\pm$0.05 & y & y & y \\
220 & 16.46 & 14.39$\pm$0.04 & 13.83$\pm$0.04 & 13.40$\pm$0.04 & y & y & y \\
221 & 16.38 & 14.41$\pm$0.04 & 13.93$\pm$0.07 & 13.48$\pm$0.05 & y & y & y \\
222 & 14.96 & 13.24$\pm$0.03 & 12.71$\pm$0.04 & 12.35$\pm$0.03 & y & y & y \\
223 & 17.15 & 15.10$\pm$0.06 & 14.61$\pm$0.08 & 14.25$\pm$0.09 & y & y & y \\
224 & 14.71 & 12.85$\pm$0.03 & 12.39$\pm$0.05 & 12.01$\pm$0.04 & y & y & y \\
225 & 15.54 & 13.71$\pm$0.03 & 13.26$\pm$0.06 & 12.91$\pm$0.03 & y & y & y \\
226 & 14.86 & 13.16$\pm$0.03 & 12.64$\pm$0.04 & 12.37$\pm$0.03 & y & y & y \\
227 & 15.44 & 13.75$\pm$0.03 & 13.27$\pm$0.05 & 12.94$\pm$0.03 & y & y & y \\
228 & 17.37 & 15.05$\pm$0.05 & 14.53$\pm$0.07 & 14.03$\pm$0.07 & y & y & y \\
229 & 15.40 & 13.58$\pm$0.03 & 13.03$\pm$0.05 & 12.66$\pm$0.04 & y & y & y \\
230 & 14.18 & 12.60$\pm$0.03 & 12.06$\pm$0.06 & 11.70$\pm$0.04 & y & y & y \\
231 & 15.10 & 13.35$\pm$0.03 & 12.85$\pm$0.03 & 12.53$\pm$0.03 & y & y & y \\
232 & 16.02 & 14.20$\pm$0.04 & 13.79$\pm$0.05 & 13.28$\pm$0.04 & y & y & y \\
233 & 13.70 & 12.27$\pm$0.03 & 11.73$\pm$0.04 & 11.39$\pm$0.03 & y & y & y \\
238 & 15.59 & 13.70$\pm$0.03 & 13.11$\pm$0.03 & 12.73$\pm$0.03 & y & y & y \\
239 & 15.19 & 13.56$\pm$0.03 & 13.06$\pm$0.03 & 12.69$\pm$0.03 & y & y & y \\
241 & 14.26 & 12.95$\pm$0.03 & 12.39$\pm$0.03 & 12.04$\pm$0.03 & y & y & y \\
242 & 16.11 & 14.26$\pm$0.03 & 13.73$\pm$0.04 & 13.29$\pm$0.04 & y & y & y \\
243 & 13.97 & 12.42$\pm$0.03 & 11.86$\pm$0.03 & 11.47$\pm$0.03 & y & y & y \\
244 & 14.29 & 12.64$\pm$0.03 & 12.11$\pm$0.03 & 11.72$\pm$0.02 & y & y & y \\
245 & 15.07 & 13.49$\pm$0.03 & 12.96$\pm$0.03 & 12.58$\pm$0.03 & y & y & y \\
246 & 15.16 & 13.60$\pm$0.03 & 13.03$\pm$0.03 & 12.74$\pm$0.04 & y & y & y \\
247 & 16.90 & 15.21$\pm$0.06 & 14.57$\pm$0.06 & 14.20$\pm$0.07 & N & y & y \\
248 & 15.16 & 13.45$\pm$0.03 & 12.95$\pm$0.03 & 12.63$\pm$0.04 & y & y & y \\
250 & 15.49 & 13.61$\pm$0.03 & 13.17$\pm$0.03 & 12.80$\pm$0.04 & y & y & y \\
251 & 15.57 & 13.84$\pm$0.03 & 13.29$\pm$0.04 & 12.95$\pm$0.04 & y & y & y \\
252 & 15.38 & 13.80$\pm$0.03 & 13.29$\pm$0.03 & 12.91$\pm$0.04 & y & y & y \\
253 & 17.24 & 15.26$\pm$0.05 & 14.63$\pm$0.06 & 14.25$\pm$0.07 & y & y & y \\
255 & 16.07 & 14.11$\pm$0.03 & 13.64$\pm$0.03 & 13.28$\pm$0.04 & y & y & y \\
260 & 15.73 & 13.97$\pm$0.03 & 13.48$\pm$0.04 & 13.07$\pm$0.04 & y & y & y \\
264 & 14.98 & 13.41$\pm$0.03 & 12.86$\pm$0.03 & 12.49$\pm$0.04 & y & y & y \\
267 & 16.41 & 14.61$\pm$0.04 & 14.06$\pm$0.05 & 13.67$\pm$0.05 & y & y & y \\
272 & 16.53 & 14.74$\pm$0.04 & 14.14$\pm$0.05 & 13.70$\pm$0.06 & y & y & y \\
276 & 14.01 & 13.17$\pm$0.03 & 12.84$\pm$0.03 & 12.70$\pm$0.04 & N & N & N \\
280 & 16.75 & 14.77$\pm$0.04 & 14.32$\pm$0.06 & 13.79$\pm$0.06 & y & y & y \\
281 & 15.79 & 14.09$\pm$0.03 & 13.55$\pm$0.04 & 13.16$\pm$0.04 & y & y & y \\
285 & 14.97 & 13.36$\pm$0.03 & 12.83$\pm$0.03 & 12.52$\pm$0.04 & y & y & y \\
290 & 14.71 & 13.14$\pm$0.03 & 12.60$\pm$0.03 & 12.27$\pm$0.03 & y & y & y \\
291 & 16.40 & 14.37$\pm$0.04 & 13.83$\pm$0.04 & 13.44$\pm$0.05 & y & y & y \\
292 & 16.73 & 14.96$\pm$0.05 & 14.46$\pm$0.06 & 14.15$\pm$0.07 & N & y & y \\
\hline
\end{tabular}
\end{center}
\end{table*}

\begin{table*}
\begin{center}
\begin{tabular}{|l|l|l|l|l|l|l|l|}
\hline
BPL & I$_C$ & J$_{MKO}$ & H$_{MKO}$ & K$_{MKO}$ & {\it IK} & {\it JK} & {\it JHK} \\
\hline
293 & 17.16 & 15.17$\pm$0.05 & 14.67$\pm$0.07 & 14.28$\pm$0.08 & y & y & y \\
296 & 16.79 & 15.07$\pm$0.05 & 14.50$\pm$0.06 & 14.11$\pm$0.07 & N & y & y \\
298 & 14.20 & 12.66$\pm$0.03 & 12.16$\pm$0.03 & 11.81$\pm$0.03 & y & y & y \\
299 & 16.61 & 14.74$\pm$0.04 & 14.24$\pm$0.05 & 13.84$\pm$0.05 & y & y & y \\
301 & 15.58 & 13.85$\pm$0.03 & 13.30$\pm$0.03 & 12.93$\pm$0.03 & y & y & y \\
305 & 14.07 & 12.62$\pm$0.03 & 12.10$\pm$0.03 & 11.74$\pm$0.02 & y & y & y \\
310 & 14.44 & 12.90$\pm$0.03 & 12.36$\pm$0.04 & 12.06$\pm$0.04 & y & y & y \\
311 & 14.96 & 13.41$\pm$0.03 & 12.85$\pm$0.03 & 12.54$\pm$0.03 & y & y & y \\
313 & 16.44 & 14.55$\pm$0.04 & 13.99$\pm$0.05 & 13.74$\pm$0.05 & y & y & y \\
314 & 13.82 & 12.65$\pm$0.03 & 12.07$\pm$0.03 & 11.78$\pm$0.03 & N & y & y \\
318 & 13.88 & 12.61$\pm$0.03 & 12.00$\pm$0.03 & 11.74$\pm$0.03 & N & y & y \\
319 & 13.75 & 12.87$\pm$0.03 & 12.44$\pm$0.03 & 12.28$\pm$0.03 & N & N & N \\
320 & 15.59 & 14.02$\pm$0.03 & 13.50$\pm$0.03 & 13.15$\pm$0.03 & y & y & y \\
321 & 14.41 & 12.91$\pm$0.03 & 12.30$\pm$0.03 & 11.93$\pm$0.04 & y & y & y \\
322 & 15.86 & 14.19$\pm$0.03 & 13.68$\pm$0.03 & 13.25$\pm$0.04 & y & y & y \\
323 & 16.27 & 14.38$\pm$0.03 & 13.93$\pm$0.04 & 13.56$\pm$0.04 & y & y & y \\
324 & 14.15 & 12.77$\pm$0.03 & 12.21$\pm$0.03 & 11.84$\pm$0.03 & y & y & y \\
325 & 17.41 & 15.21$\pm$0.05 & 14.81$\pm$0.07 & 14.28$\pm$0.08 & y & y & y \\
326 & 16.96 & 15.06$\pm$0.04 & 14.50$\pm$0.05 & 14.27$\pm$0.08 & N & y & y \\
327 & 18.00 & 15.52$\pm$0.05 & 14.93$\pm$0.07 & 14.78$\pm$0.12 & y & y & y \\
328 & 16.95 & 14.72$\pm$0.04 & 14.10$\pm$0.04 & 13.68$\pm$0.05 & y & y & y \\
329 & 14.52 & 13.05$\pm$0.03 & 12.47$\pm$0.03 & 12.15$\pm$0.03 & y & y & y \\
330 & 15.73 & 14.16$\pm$0.03 & 13.57$\pm$0.03 & 13.23$\pm$0.04 & y & y & y \\
331 & 17.67 & 15.41$\pm$0.05 & 14.75$\pm$0.06 & 14.34$\pm$0.09 & y & y & y \\
332 & 16.41 & 14.59$\pm$0.04 & 14.04$\pm$0.05 & 13.66$\pm$0.05 & y & y & y \\
333 & 14.69 & 13.02$\pm$0.03 & 12.45$\pm$0.03 & 12.13$\pm$0.03 & y & y & y \\
334 & 17.47 & 15.27$\pm$0.05 & 14.67$\pm$0.06 & 14.31$\pm$0.08 & y & y & y \\
335 & 15.36 & 13.55$\pm$0.03 & 13.05$\pm$0.04 & 12.66$\pm$0.03 & y & y & y \\
336 & 17.46 & 15.30$\pm$0.05 & 14.70$\pm$0.06 & 14.26$\pm$0.08 & y & y & y \\
337 & 16.71 & 14.69$\pm$0.04 & 14.15$\pm$0.04 & 13.73$\pm$0.05 & y & y & y \\
338 & 16.81 & 15.07$\pm$0.04 & 14.52$\pm$0.05 & 14.24$\pm$0.07 & N & y & y \\
339 & 13.84 & 12.43$\pm$0.03 & 11.88$\pm$0.03 & 11.46$\pm$0.03 & y & y & y \\
\hline
\end{tabular}
\end{center}
\end{table*}

\begin{table*}
\begin{center}
\caption{Photometry (c2-5), membership (c6-8) and binarity (c9) of RIZ-Pr Praesepe candidates. 
Membership is from the IK CMD, the JK CMD and the JHK diagram. Refs; (a) Pinfield et~al. (1997), 
(b) Hodgkin et al. (1999). The $^{\dagger}$ is also IZ-Pr 40. The $^{\ddagger}$ is also IZ-Pr 86. 
n-d = non-detection}
\begin{tabular}{|l|l|l|l|l|l|l|l|l|l|l|l|}
\hline
RIZ-Pr & I$_C^a$ & J$_{MKO}$ & H$_{MKO}$ & K$_{MKO}$ & {\it IK} & {\it JK} & {\it JHK} & Bin \\
\hline
 1 & 20.14 & 17.06$\pm$0.05   & 16.47$\pm$0.04   & 16.14$\pm$0.08$^b$ & y & y & y  &   \\
 2$^{\dagger}$ & 18.19 & 15.67$\pm$0.05 & 15.11$\pm$0.03 & 14.77$\pm$0.04$^b$ & y & y & y  & b \\
 3 & 21.20 & 19.05$\pm$0.17   & 18.01$\pm$0.12   & 16.77$\pm$0.08     & y & N & N  &   \\
 4 & 18.52 & 16.04$\pm$0.03   & 15.56$\pm$0.03   & 15.23$\pm$0.04$^b$ & y & y & y  &   \\
 5 & 20.86 & 19.09$\pm$0.17   & 18.20$\pm$0.16   & 16.94$\pm$0.06     & y & N & N  &   \\
 6 & 21.09 & 19.10$\pm$0.18   & 18.11$\pm$0.16   & 17.07$\pm$0.08     & y & N & N  &   \\
   &       &                  &                  & 17.01$\pm$0.07     &   &   &    &   \\
   &       &                  &                  & 17.32$\pm$0.05$^b$ &   &   &    &   \\
 7 & 21.24 & --               & --               & 18.56$\pm$0.29     & N & --& -- &   \\
   &       &                  &                  & 19.15$\pm$0.05$^b$ &   &   &    &   \\
 8 & 17.81 & 15.40$\pm$0.05   & 14.89$\pm$0.03   & 14.60$\pm$0.04$^b$ & y & y & y  &   \\
 9 & 21.25 & --               & --               & 17.72$\pm$0.13     & N & --& -- &   \\
10 & 20.54 & --               & --               & 18.08$\pm$0.07$^b$ & N & --& -- &   \\
11 & 19.47 & 16.66$\pm$0.05   & 16.23$\pm$0.04   & 15.84$\pm$0.14$^b$ & y & y & y  &   \\
12 & 20.10 & --               & --               & 16.91$\pm$0.21$^b$ & N & --& -- &   \\
13 & 20.24 & --               & --               & 19.04$\pm$0.50     & N & --& -- &   \\
   &       &                  &                  & 18.49$\pm$0.11$^b$ &   &   &    &   \\
14 & 20.59 & --               & --               & 17.82$\pm$0.06$^b$ & N & --& -- &   \\
15 & 21.36 & --               & --               & 18.49$\pm$0.27     & N & --& -- &   \\
16 & 21.10 & --               & --               & 18.60$\pm$0.31     & N & --& -- &   \\
17 & 20.41 & 18.23$\pm$0.10   & 17.65$\pm$0.12   & 17.28$\pm$0.11     & N & N & y  &   \\
18 & 19.63 & 16.40$\pm$0.04   & 15.74$\pm$0.03   & 15.40$\pm$0.04$^b$ & y & y & N? & b \\
19 & 21.42 & --               & --               & 18.04$\pm$0.18     & N & --& -- &   \\
20 & 18.48 & 16.16$\pm$0.04   & 15.57$\pm$0.03   & 15.19$\pm$0.04$^b$ & y & y & y  &   \\
21$^{\ddagger}$ & 18.73 & 15.89$\pm$0.03 & 15.39$\pm$0.03 & 15.03$\pm$0.03$^b$ & y & y & y & b \\
22 & 21.55 & --               & --               & n-d                & N & --& -- &   \\
23 & 19.06 & 16.63$\pm$0.04   & 16.04$\pm$0.03   & 15.53$\pm$0.03     & y & y & y & b \\
   &       &                  &                  & 15.98$\pm$0.16$^b$ &   &   &    &   \\
24 & 20.43 & 17.30$\pm$0.05   & 16.68$\pm$0.04   & 16.17$\pm$0.06     & y & y & y  & b \\
   &       &                  &                  & 16.19$\pm$0.03$^b$ &   &   &    &   \\
25 & 21.20 & --               & --               & 18.62$\pm$0.31     & N & --& -- &   \\
26 & 18.08 & --               & --               & 15.49$\pm$0.04$^b$ & N & --& -- &   \\
\hline
\end{tabular}
\end{center}
\end{table*}

\begin{table*}
\begin{center}
\caption{Photometry (c2-5), membership (c6-8) and binarity (c9) of IZ-Pr Praesepe candidates. 
Membership is from the IK CMD, the JK CMD and the JHK diagram. Refs; (a) Pinfield etal (2000), 
(b) 2MASS. The $^{\dagger}$ is also RIZ-Pr 2. The $^{\ddagger}$ is also RIZ-PR 21.}
\begin{tabular}{|l|l|l|l|l|l|l|l|l|}
\hline
IZ-Pr & I$_C^a$ & J$_{MKO}$ & H$_{MKO}$ & K$_{MKO}$ & {\it IK} & {\it JK} & {\it JHK} & Bin \\
\hline
  1 & 18.02 & 15.60$\pm$0.03     & 15.13$\pm$0.05     & 14.72$\pm$0.03     & y & y  & y  & b \\
  2 & 17.25 & 15.44$\pm$0.03     & 14.94$\pm$0.05     & 14.64$\pm$0.03     & y & y  & y  &   \\
  3 & 18.19 & 15.87$\pm$0.03     & 15.36$\pm$0.05     & 15.09$\pm$0.02     & y & y  & y  &   \\
  4 & 18.21 & 16.56$\pm$0.13$^b$ & 15.76$\pm$0.13$^b$ & 15.31$\pm$0.06     & y & y  & ?  &   \\
  5 & 18.33 & --                 & --                 & 15.83$\pm$0.03     & N & -- & -- &   \\
  6 & 17.26 & 15.62$\pm$0.06$^b$ & 15.15$\pm$0.08$^b$ & 14.79$\pm$0.03     & y & y? & ?  &   \\
  7 & 19.90 & --                 & --                 & 16.52$\pm$0.05     & N & -- & -- &   \\
  8 & 17.22 & 15.57$\pm$0.06$^b$ & 15.16$\pm$0.08$^b$ & 14.72$\pm$0.03     & y & y? & ?  &   \\
  9 & 20.37 & --                 & --                 & 17.66$\pm$0.12     & N & -- & -- &   \\
 10 & 18.26 & 15.79$\pm$0.03     & 15.30$\pm$0.05     & 14.95$\pm$0.02     & y & y  & y  & b \\
 11 & 17.47 & 15.63$\pm$0.03     & 15.10$\pm$0.05     & 14.82$\pm$0.03     & y & y  & y  &   \\
 12 & 19.93 & --                 & --                 & 17.36$\pm$0.09     & N & -- & -- &   \\
 13 & 19.14 & --                 & --                 & 16.08$\pm$0.04     & N & -- & -- &   \\
 14 & 17.84 & 16.03$\pm$0.07$^b$ & 15.50$\pm$0.09$^b$ & 15.02$\pm$0.03     & y & y  & ?  &   \\
 15 & 17.94 & 16.08$\pm$0.08$^b$ & 15.63$\pm$0.09$^b$ & 15.11$\pm$0.03     & y & y  & ?  &   \\
 16 & 17.18 & --                 & --                 & 14.99$\pm$0.03     & N & -- & -- &   \\
 17 & 17.61 & 15.74$\pm$0.03     & 15.24$\pm$0.05     & 14.86$\pm$0.03     & y & y  & y  &   \\
 18 & 18.54 & 16.18$\pm$0.03     & 15.81$\pm$0.03     & 15.30$\pm$0.04     & y & y  & N  &   \\
 19 & 19.93 & --                 & --                 & 16.78$\pm$0.06     & N & -- & -- &   \\
 20 & 17.22 & 15.43$\pm$0.03     & 14.92$\pm$0.05     & 14.58$\pm$0.03     & y & y  & y  &   \\
 21 & 19.72 & --                 & --                 & 17.19$\pm$0.08     & N & -- & -- &   \\
 22 & 19.65 & --                 & --                 & 16.18$\pm$0.04     & y & -- & -- &   \\
 23 & 19.20 & 16.47$\pm$0.03     & 16.04$\pm$0.03     & 15.57$\pm$0.04     & y & y  & y  &   \\
 24 & 17.52 & 15.47$\pm$0.03     & 14.95$\pm$0.05     & 14.56$\pm$0.03     & y & y  & y  & b \\
 25 & 17.98 & 15.49$\pm$0.03     & 14.94$\pm$0.05     & 14.64$\pm$0.02     & y & y  & y  & b \\
 26 & 17.38 & 15.30$\pm$0.03     & 14.86$\pm$0.04     & 14.52$\pm$0.03     & y & y  & y  &   \\
 27 & 19.81 & --                 & --                 & 16.77$\pm$0.06     & N & -- & -- &   \\
 28 & 17.89 & 16.22$\pm$0.09$^b$ & 15.50$\pm$0.09$^b$ & 15.15$\pm$0.03     & y & y  & ?  &   \\
 29 & 19.62 & --                 & --                 & 16.90$\pm$0.06     & N & -- & -- &   \\
 30 & 17.55 & 15.51$\pm$0.03     & 15.09$\pm$0.05     & 14.72$\pm$0.03     & y & y  & y  &   \\
 31 & 17.41 & 15.38$\pm$0.03     & 14.91$\pm$0.02     & 14.54$\pm$0.03     & y & y  & y  &   \\
 32 & 19.19 & --                 & --                 & 15.88$\pm$0.03     & y & ?  & -- &   \\
 33 & 17.30 & 15.36$\pm$0.03     & 14.90$\pm$0.05     & 14.55$\pm$0.03     & y & y  & y  &   \\
 34 & 19.34 & --                 & --                 & 17.49$\pm$0.10     & N & -- & -- &   \\
 35 & 19.77 & --                 & --                 & 16.50$\pm$0.05     & N & -- & -- &   \\
 36 & 19.41 & 16.62$\pm$0.04     & 16.11$\pm$0.03     & 15.66$\pm$0.02     & y & y  & y  &   \\
 37 & 19.28 & --                 & --                 & 16.40$\pm$0.04     & N & -- & -- &   \\
 38 & 18.22 & --                 & --                 & 15.52$\pm$0.03     & N & -- & -- &   \\
 39 & 19.44 & --                 & --                 & 16.58$\pm$0.05     & N & -- & -- &   \\
 40$^{\dagger}$ & 18.04 & 15.74$\pm$0.04     & 15.08$\pm$0.03     & 14.79$\pm$0.04     & y & y  & y  &   \\
 41 & 18.02 & 15.75$\pm$0.04     & 15.28$\pm$0.02     & 14.83$\pm$0.03     & y & y  & y  & b \\
 42 & 20.33 & --                 & --                 & 17.03$\pm$0.08     & N & -- & -- &   \\
 43 & 19.47 & --                 & --                 & 16.09$\pm$0.10     & y & -- & -- &   \\
 44 & 17.49 & 15.76$\pm$0.06$^b$ & 15.32$\pm$0.08$^b$ & 14.84$\pm$0.04     & y & y  & ?  &   \\
 45 & 17.18 & --                 & --                 & 15.46$\pm$0.11     & N & -- & -- &   \\
 50 & 18.29 & --                 & --                 & 15.43$\pm$0.03     & y & -- & -- &   \\
 52 & 19.21 & --                 & --                 & 16.27$\pm$0.04     & N & -- & -- &   \\
 54 & 17.67 & 15.87$\pm$0.08$^b$ & 15.39$\pm$0.10$^b$ & 15.07$\pm$0.14$^b$ & y & y? & ?  &   \\
 55 & 17.40 & 15.74$\pm$0.07$^b$ & 15.18$\pm$0.08$^b$ & 15.15$\pm$0.15$^b$ & N & N  & ?  &   \\
 56 & 17.38 & 15.51$\pm$0.06$^b$ & 14.93$\pm$0.07$^b$ & 14.53$\pm$0.09$^b$ & y & y  & ?  &   \\
 57 & 17.45 & 15.62$\pm$0.06$^b$ & 15.01$\pm$0.07$^b$ & 14.62$\pm$0.09$^b$ & y & y  & ?  &   \\
 59 & 18.47 & --                 & --                 & 15.83$\pm$0.04     & N & -- & -- &   \\
 60 & 17.48 & 15.60$\pm$0.06$^b$ & 15.20$\pm$0.09$^b$ & 14.81$\pm$0.11$^b$ & y & y? & ?  &   \\
 61 & 17.59 & 16.07$\pm$0.09$^b$ & 15.38$\pm$0.10$^b$ & 14.99$\pm$0.13$^b$ & y & y  & ?  &   \\
 62 & 17.77 & 16.07$\pm$0.09$^b$ & 15.55$\pm$0.11$^b$ & 15.11$\pm$0.14$^b$ & y & y  & ?  &   \\
 63 & 17.42 & 15.72$\pm$0.06$^b$ & 14.99$\pm$0.08$^b$ & 14.51$\pm$0.08$^b$ & y & y  & ?  & b \\
 64 & 17.44 & 15.30$\pm$0.05$^b$ & 14.66$\pm$0.06$^b$ & 14.19$\pm$0.06$^b$ & y & y  & ?  & b \\
 66 & 17.26 & 15.40$\pm$0.05$^b$ & 14.81$\pm$0.07$^b$ & 14.61$\pm$0.09$^b$ & y & y? & ?  &   \\
 69 & 17.30 & 15.45$\pm$0.05$^b$ & 14.89$\pm$0.07$^b$ & 14.53$\pm$0.09$^b$ & y & y  & ?  &   \\
\hline
\end{tabular}
\end{center}
\end{table*}

\begin{table*}
\begin{center}
\begin{tabular}{|l|l|l|l|l|l|l|l|l|}
\hline
IZ-Pr & I$_C^a$ & J$_{MKO}$ & H$_{MKO}$ & K$_{MKO}$ & {\it IK} & {\it JK} & {\it JHK} & Bin \\
\hline
 70 & 17.32 & 15.40$\pm$0.05$^b$ & 14.74$\pm$0.07$^b$ & 14.53$\pm$0.09$^b$ & y & y? & ?  &   \\
 72 & 17.29 & 15.38$\pm$0.05$^b$ & 14.93$\pm$0.07$^b$ & 14.36$\pm$0.09$^b$ & y & y  & ?  & b \\
 74 & 17.87 & 15.89$\pm$0.07$^b$ & 15.06$\pm$0.08$^b$ & 14.90$\pm$0.12$^b$ & y & y  & ?  &   \\
 77 & 17.41 & 15.71$\pm$0.06$^b$ & 15.14$\pm$0.08$^b$ & 14.76$\pm$0.10$^b$ & y & y  & ?  &   \\
 80 & 17.37 & 15.55$\pm$0.05$^b$ & 14.81$\pm$0.07$^b$ & 14.52$\pm$0.08$^b$ & y & y  & ?  &   \\
 82 & 19.24 & --                 & --                 & 16.54$\pm$0.04     & N & -- & -- &   \\
 84 & 17.99 & 15.95$\pm$0.07$^b$ & 15.26$\pm$0.08$^b$ & 15.13$\pm$0.14$^b$ & y & y? & ?  &   \\
 85 & 18.92 & --                 & --                 & 17.03$\pm$0.05     & N & -- & -- &   \\
 86$^{\ddagger}$ & 18.63 & 15.89$\pm$0.03     & 15.39$\pm$0.03     & 15.01$\pm$0.03     & y & y  & y  &   \\
 88 & 17.98 & 16.11$\pm$0.09$^b$ & 15.85$\pm$0.15$^b$ & 15.15$\pm$0.15$^b$ & y & y  & ?  &   \\
 93 & 19.29 & 16.76$\pm$0.04     & 16.21$\pm$0.03     & 15.75$\pm$0.03     & y & y  & y  &   \\
 94 & 18.25 & 16.02$\pm$0.10$^b$ & 15.44$\pm$0.11$^b$ & 15.00$\pm$0.03     & y & y  & -- & b \\
 95 & 19.47 & --                 & --                 & 16.78$\pm$0.03     & N & -- & -- &   \\
 97 & 19.50 & --                 & --                 & 16.75$\pm$0.04     & N & -- & -- &   \\
 99 & 17.86 & 15.88$\pm$0.09$^b$ & 15.19$\pm$0.10$^b$ & 14.98$\pm$0.13$^b$ & y & y  & ?  &   \\
100 & 17.34 & 15.40$\pm$0.06$^b$ & 14.95$\pm$0.09$^b$ & 14.59$\pm$0.08$^b$ & y & y? & ?  &   \\
104 & 19.50 & --                 & --                 & 16.84$\pm$0.06     & N & -- & -- &   \\
105 & 18.80 & --                 & --                 & 16.15$\pm$0.03     & N & -- & -- &   \\
108 & 17.31 & 15.47$\pm$0.06$^b$ & 14.83$\pm$0.07$^b$ & 14.57$\pm$0.10$^b$ & y & y  & ?  &   \\
110 & 18.68 & --                 & --                 & 16.74$\pm$0.05     & N & -- & -- &   \\
112 & 17.89 & 16.01$\pm$0.09$^b$ & 15.53$\pm$0.12$^b$ & 15.17$\pm$0.15$^b$ & y & y? & ?  &   \\
113 & 18.44 & 16.03$\pm$0.03     & 15.58$\pm$0.03     & 15.17$\pm$0.03     & y & y  & y  &   \\
117 & 19.09 & 16.56$\pm$0.04     & 16.02$\pm$0.03     & 15.70$\pm$0.03     & y & y  & y  &   \\
119 & 17.94 & 15.77$\pm$0.08$^b$ & 15.00$\pm$0.08$^b$ & 14.66$\pm$0.09$^b$ & y & y  & ?  & b \\
120 & 17.48 & 15.68$\pm$0.07$^b$ & 15.28$\pm$0.09$^b$ & 14.75$\pm$0.11$^b$ & y & y  & ?  &   \\
123 & 19.49 & 15.79$\pm$0.08$^b$ & 15.41$\pm$0.11$^b$ & 15.26$\pm$0.15$^b$ & y & N  & ?  &   \\
125 & 17.48 & 16.01$\pm$0.09$^b$ & 15.52$\pm$0.11$^b$ & 15.21$\pm$0.15$^b$ & N & y? & ?  &   \\
126 & 18.79 & 15.94$\pm$0.09$^b$ & 15.57$\pm$0.12$^b$ & 15.22$\pm$0.02     & y & y? & ?  & b \\
127 & 19.98 & 16.48$\pm$0.16$^b$ & 16.18$\pm$0.23$^b$ & 15.49$\pm$0.19$^b$ & y & y  & ?  &   \\
128 & 17.50 & 15.57$\pm$0.07$^b$ & 14.89$\pm$0.08$^b$ & 14.62$\pm$0.09$^b$ & y & y  & ?  &   \\
\hline
\end{tabular}
\end{center}
\end{table*}

\label{lastpage}


\begin{thebibliography}{99}
\bibitem{} Allard F., Hauschildt P. H., Alexander D. R., Starrfield S., 1997, ARA\& A, 35, 137
\bibitem{} Allard F., Hauschildt P. H., Alexander D. R., Tamanai A., Schweitzer A., 2001, 
   ApJ, 556, 357
\bibitem{} Bailer-Jones C. A. L., Mundt R., 2001, A\& A, 367, 218
\bibitem{} Baraffe I., Chabrier G., Allard F., Hauschildt P. H., 1998, A\& A, 337, 403
\bibitem{} Barrado y Navascu\'es D., Bouvier J., Stauffer J. R., Lodieu N., McCaughrean M. J., 
   2002, A\& A, 395, 813
\bibitem{} Basri G., Mart\'{\i}n E. L., 1999, AJ, 118, 2460
\bibitem{} Bate M. R., 1998, ApJ, 508, L95
\bibitem{} Bate M. R., Bonnell I. A., Bromm V., 2002, MNRAS, 332, 65
\bibitem{} Bessell M. S., 1986, PASP, 98, 1303
\bibitem{} Bonnell I. A., Bate M. R., 1994, MNRAS, 271, 999
\bibitem{} Bouvier J., Stauffer J. R., Mart\'{\i}n E. L., Barrado y Navascues D., Wallace B., 
   Bejar V. J. S., 1998, A\& A 336, 490
\bibitem{} Burgasser A. J., Kirkpatrick J. D., Reid I. N., Brown M. E., Miskey C. L., Gizis J. E., 
   2003, ApJ, in press
\bibitem{} Burrows A., Hubbard W. B., Saumon D., Lunine J. I., 1993, ApJ, 406, 158
\bibitem{} Burrows A., Marley M., Hubbard W. B., Lunine J. I., Guillot T., Saumon D., Freedman R., 
   Sudarsky D., Sharp C., 1997, ApJ, 491, 856
\bibitem{} Carpenter J. M., 2001, AJ, 121, 2851
\bibitem{} Chabrier G., Baraffe I., Allard F., Hauschildt P. H., 2000, ApJ, 542, 464
\bibitem{} Chabrier G., Baraffe I., 1997, A\& A, 327, 1039
\bibitem{} Chappelle R. J., Pinfield D. J., Steele I. A., 2002, in proceedings of the IAU 
   Symposium 211 on Brown Dwarfs, in press
\bibitem{} Close L. M., Siegler N., Freed M., Biller B., 2003, ApJ, 587, 407
\bibitem{} Crawford D. L., Perry C. L., 1976, AJ, 81, 419
\bibitem{} Dahn C. C., Harris H. C., Vrba F. J., Guetter H. H., Canzian B., Henden A. A., 
   Levine S. E., Luginbuhl C. B., Monet A. K. B., Monet D. G., Pier J. R., Stone R. C., 
   Walker R. L., Burgasser A. J., Gizis J. E., Kirkpatrick J. D., Leibert J., Reid I. N., 
   2002, AJ, 124, 1170
\bibitem{} D'Antona F., Mazzitelli I., 1994, ApJS, 90, 467
\bibitem{} Dobbie P. D., Kenyon F., Jameson R. F., Hodgkin S. T., 2002a, MNRAS, 331, 445
\bibitem{} Dobbie P. D., Kenyon F., Jameson R. F., Hodgkin S. T., Pinfield D. J., Osbourne 
   S. L., 2002b, MNRAS, 335, 687
\bibitem{} Dobbie P. D., Pinfield D. J., Jameson R. F., Hodgkin S. T., 2002c, MNRAS, 335, 79L
\bibitem{} Duquennoy A., Mayor M., 1991, A\& A, 248, 485
\bibitem{} Friel E. D., Boesgaard A. M., 1992, ApJ, 387, 170
\bibitem{} Hambly N. C., Hawkins M. R. S., Jameson R. F., 1993, A\& AS, 100, 607
\bibitem{} Hambly N. C., Steele I. A., Hawkins M. R. S., Jameson R. F., 1995a, A\& AS, 
   109, 29 (HSHJ)
\bibitem{} Hambly N. C., Steele I. A., Hawkins M. R. S., Jameson R. F., 1995b, MNRAS, 273, 505
\bibitem{} Hauschildt P. H., Allard F., Baron E., 1999, ApJ, 512, 377
\bibitem{} Hawarden T. G., Leggett S. K., Letawsky M. B., Ballantyne D. R., Casali M. M., 
   2001, MNRAS, 325, 563
\bibitem{} Hawley S. L., Tourtellot J. G., Reid I. N., 1999, AJ, 117, 1341
\bibitem{} Hodgkin S. T., Pinfield D. J., Jameson R. F., Steele I. A., Cossburn M. R., 
   Hambly N. C., 1999, MNRAS, 310, 87
\bibitem{} Holland K., Jameson R. F., Hodgkin S. T., Davies M. B., Pinfield D. J., 2000, 
   MNRAS, 319, 956
\bibitem{} Jameson R. F., Dobbie P. D., Hodgkin S. T., Pinfield D. J., 2002, MNRAS, 335, 853
\bibitem{} Jeffries R. D., Thurston M. R., Hambly N. C., 2001, A\& A, 375, 863
\bibitem{} Jones H. R. A., Tsuji T., 1997, ApJ, 480, L39
\bibitem{} King I. R., 1962, AJ, 67, 471
\bibitem{} Kirkpatrick J. D., Henry T. J., McCarthy D. W. Jr., 1991, ApJS, 77, 417
\bibitem{} Kirkpatrick J. D., Henry T. J., Simons D. A., 1995, AJ, 109, 797
\bibitem{} Kirkpatrick J. D., McGraw J. T., Hess T. R., Liebert J., McCarthy D. W. Jr., 
   1994, ApJS, 94, 749
\bibitem{} Kroupa P., 2001, MNRAS, 322, 231
\bibitem{} Kroupa P., Aarseth S., Hurley J., 2001, MNRAS, 321, 699
\bibitem{} Kroupa P., Tout C. A., Gilmore G., 1990, MNRAS, 244, 76
\bibitem{} Leggett S. K., 1992, ApJS, 82, 351
\bibitem{} Leggett S. K., Allard F., Hauschildt P. H., 1998, ApJ, 509, 836
\bibitem{} Liebert J., Kirkpatrick J. D., Reid I. N., Fisher M. D., 1999, ApJ, 519, 345
\bibitem{} Lockwood G. W., Thompson D. T., Radick R. R., Osborn W. H., Baggett W. E., 
   Duncan D. K., Hartmann L. W., 1984, PASP, 96, 714
\bibitem{} Lucas P. W., Roche P. F., 2000, MNRAS, 314, 858
\bibitem{} Lynga G., 1987, Catalogue, 5th edition (Strasbourg)
\bibitem{} Magazzu A., Rebolo R., Zapatero-Osorio M. R., Mart\'{\i}n E. L., Hodgkin S. T., 
   1998, ApJ, 497, 47
\bibitem{} Makarov, V. V., 2002, AJ, 124, 3299
\bibitem{} Mart\'{\i}n E. L., Basri G., Brandner W., Bouvier J., Zapatero-Osorio M. R., 
   Rebolo R., Stauffer J. R., Allard F., Baraffe I., Hodgkin S. T., 1998, ApJ, 509, 113
\bibitem{} Mart\'{\i}n E. L., Brandner W., Bouvier J., Luhman K. L., Stauffer J., Basri G., 
   Zapatero-Osorio M. R., Barrado y Navascues D., 2000, ApJ, 543, 299
\bibitem{} Mart\'{\i}n E. L., Delfosse X., Basri G., Goldman B., Forveille T., Zapatero-Osorio 
   M. R., 1999, AJ, 118, 2466
\bibitem{} Mart\'{\i}n E. L., Kun M., 1996, A\& AS 116, 467
\bibitem{} Mart\'{\i}n E. L., Rebolo R., Zapatero-Osorio M. R., 1996, ApJ, 469, 706
\bibitem{} Mermilliod J. -C., Turon C., Robichon N., Arenou F., Lebreton Y, 1997, in proceedings of 
the ESA Symposium `Hipparcos - Venice '97', 643
\bibitem{} Nissen P. E., 1988, A\& A, 199, 146
\bibitem{} Pinfield D. J., 1997, PhD Thesis, Univ. Leicester
\bibitem{} Pinfield D. J., Hodgkin S. T., Jameson R. F., Cossburn M. R., Hambly N. C., 
   Devereux N., 2000, MNRAS, 313, 347
\bibitem{} Pinfield D. J., Hodgkin S. T., Jameson R. F., Cossburn M. R., von Hippel T., 
   1997, MNRAS, 287, 180
\bibitem{} Pinfield D. J., Jameson R. F., Hodgkin S. T., 1998, MNRAS, 299, 955
\bibitem{} Prosser C. F., 1992, AJ, 103, 488
\bibitem{} Reglero V., Fabregat J., 1991, A\& AS, 90, 25
\bibitem{} Reid I. N., Cruz K. L., 2002, AJ, 123, 2806
\bibitem{} Reipurth B., Clarke C., 2001, AJ, 122, 432
\bibitem{} Scholz A., Eisloffel J., Mundt R., 2000, Astronomische Gesellschaft Abstract Series, 
   17, A05
\bibitem{} Scrutskie M. F., Beichman C., Capps R., Carpenter J., Chester T., Cutri R., Elias J., 
   Elston R., Huchra J., Liebert J., Lonsdale C., Monet D., Price S., Schneider S., Seitzer P., 
   Stiening R., Strom S., Weinberg M., 1995, AAS, 187, 7507
\bibitem{} Sharp C. M., Huebner W. F., 1990, ApJS, 72, 417
\bibitem{} Simons D. A., Tokunaga A., 2002, PASP, 114, 169
\bibitem{} Stauffer J. R., Barrado y Navascu\'es D., Bouvier J., Morrison H. L., Harding P., 
   Luhman K. L., Stanke T., McCaughrean M., Terndrup D. M., Allen L., Assoud P, 1999, ApJ, 527, 219
\bibitem{} Stauffer J. R., Giampapa M. S., Herbst W., Vincent J. M., Hartmann L. W., Stern R. A., 
   1991, ApJ, 374, 142
\bibitem{} Stauffer J. R., Schultz G., Kirkpatrick J. D., 1998, ApJ, 499, L199
\bibitem{} Tsuji T., 2003, ApJ, in press (astro-ph/0204401)
\bibitem{} van Leeuwen F., 1999, A\& A, 341, 71L
\bibitem{} Zapatero-Osorio M. R., Rebolo R., Mart\'{\i}n E. L., Hodgkin S. T., Cossburn M. R., 
   Magazzu A., Steele I. A., Jameson R. F., 1999, A\& A Supp., 134, 537

\end{thebibliography}
\end{document}